\documentclass[useAMS,usenatbib]{mn2e}

\def\rt{r_{\rm t}}
\def\rh{r_{\rm h}}
\def\st{s_{\rm t}}
\def\rp{r_{\rm p}}
\def\vp{v_{\rm p}}
\def\Ep{E_{\rm p}}
\def\Jp{J_{\rm p}}
\def\Js{J_{\rm s}}
\def\jp{j_{\rm p}}
\def\rs{r_{\rm s}}
\def\vs{v_{\rm s}}

\def\aap{AA}
\def\apjl{ApJL}
\def\mnras{MNRAS}
\def\apj{ApJ}
\def\aj{AJ}

\def\nat{Nat}

\usepackage{graphicx}
\usepackage{float}
\usepackage{bm} 
\usepackage{amssymb}
\usepackage{amsmath} 
\usepackage[export]{adjustbox}
\usepackage{mathrsfs} 
\usepackage{mathtools}

\title[Tidal streams across the mass scale]{On feathers, bifurcations and shells: the dynamics of tidal streams across the mass scale}
\author[N. C. Amorisco]{N. C. Amorisco$^{1}$\thanks{E-mail:
amorisco@dark-cosmology.dk} \\
$^{1}$Dark Cosmology Centre, Niels Bohr Institute, University of Copenhagen, Juliane Maries Vej 30, DK-2100, Copenhagen}

\begin{document}



\maketitle

\label{firstpage}

\begin{abstract}
I present an organic description of the spectrum of regimes of collisionless tidal streams and define 
the orderings between the relevant physical quantities that shape their morphology.
Three fundamental dichotomies are identified and described in the form of dimensionless inequalities. 
These govern (i) the speed of the stream's growth, (ii) the internal coherence of the stream, (iii) its thickness or opening angle, within and outside the orbital plane. 
The mechanisms through which such main qualitative properties are regulated
{ and the relevant limiting cases} are analysed. For example, the slope of the host's density profile strongly influences the
speed of the stream's growth, in both length and width, as steeper density profiles enhance differential streaming.
Internal coherence is the natural requirement for the appearance of substructure and overdensities in tidal debris, 
and I concentrate on the characteristic `feathering' typical of streams of star clusters. Overdensities and substructures
are associated with minima in the relative streaming velocity of the stream members. For streams with high circularity, these
are caused by the epicyclic oscillations of stars; however, for highly non-circular progenitor's orbits, internal substructure 
is caused by the oscillating differences in energy and actions with which material is shed at different orbital phases of the progenitor.
This modulation results in different streaming speeds along the tidal arm: the streakline of material shed between two successive 
apocentric passages is folded along its length, pulled at its centre by the faster differential streaming of particles released near pericenter,
which are therefore more widely scattered. 
When the stream is coherent enough, the same mechanism is potentially capable of generating a bimodal profile in the density 
distributions of the longer wraps of more massive progenitors, which I dub `bifurcations'. The conditions that allow streams to be internally coherent 
are explored and I comment on the cases of Palomar~5, Willman~1, the Anticenter and Sagittarius' streams. Analytical methods are 
accompanied by numerical experiments, performed using a purposely built generative model, also presented here.

\end{abstract}


\section{Introduction}

In a universe in which galaxies grow by continuously accreting and dissolving star clusters and smaller galaxies, the study of tidal features 
provides a natural and powerful tool to tackle a variety of issues: 
(i) reconstruct the assembly histories of galaxies and haloes \citep[e.g., with decreasing mass,][]{AM12, RA12, Co13, FC14, BG10, DJ12, Ve14, MD12, AN14a}; 
(ii) measure their mass profiles and density slopes out to radii that is very difficult to probe by other means \citep[e.g.,][]{GS14, Ki14}; 
(iii) map their detailed structural properties, like tridimensional shape and orientation \citep[e.g.,][]{BJ08, KS10, La10, Va11, PJ12, De13, Lu13, VC13, SJ13b, Kh14}. 

The process of tidal disruption is a natural product of the laws of gravity: smaller, bound objects are often formed within or 
captured by the deeper gravitational well of a more massive system. Given a dense enough host, such satellites remain 
bound for a finite amount of time only, and eventually end their lives shredded by the tidal forces. Although with a diverse range of 
dynamical regimes, phenomenologies, and timescales, this is a very common fate across extremely different mass scales. 
In this paper, I address the mechanisms that govern the dynamics of the formation, growth and dispersal of tidal features
in collisionless systems. With increasing masses, this includes: the formation of thin stellar streams from the slow evaporation of star clusters;
the wrapping of tidal tails from disrupting dwarf galaxies within the haloes of $L_*$ galaxies like the Milky Way; the phase 
mixing of the shells often formed by galaxies accreted onto massive ellipticals and bright centrals.

The largest and highest quality datasets on streams and substructures of tidal origin pertain the Milky Way stellar halo, 
which is scarred by the traces of disrupting globular clusters \citep[GCs, e.g.,][]{Od01, BV06a, Gr06a, Gr06b} and dwarf 
galaxies \citep[e.g.,][]{BV06b, BV14, Gr06c, KS12, MS03, Ne09}. Tremendous prospects in this field are promised 
by the upcoming precision astrometry that the now ongoing GAIA mission \citep[see e.g.,][]{Pe01, BJ13} will deliver in the coming years.
This has sparked significant activity on the subject of tidal features, although with a narrowing of interests towards the Galactic environment, 
and with a focussing of both dynamical studies and modelling techniques \citep[e.g.,][]{SJ13a, BJ14, PW14, SJ14, SR14}. 

Here, I present an organic description of the different regimes of collisionless tidal streams 
and identify the orderings between the relevant physical quantities that define and shape them.
This work builds on analyses by \citet{JK98} and \citet{JK01}, which have previously grasped an 
understanding of the physical mechanisms and timescales associated with the main properties of a 
tidal streamer, such as its width and length. These works have concentrated 
in particular on the tidal features of dwarf galaxies orbiting a MW-like host \citep[see also][]{JK08}. 
In this paper I generalise the mentioned studies by widening the progenitors' mass spectrum
and by systematically highlighting the role of relevant physical ingredients. These include quantities 
whose effect has perhaps been under appreciated, as the slope of the host's density profile, 
the progenitor's internal kinematics and ordered rotation, the details of the shedding history.

\begin{figure*}
\centering
\includegraphics[width=.328\textwidth]{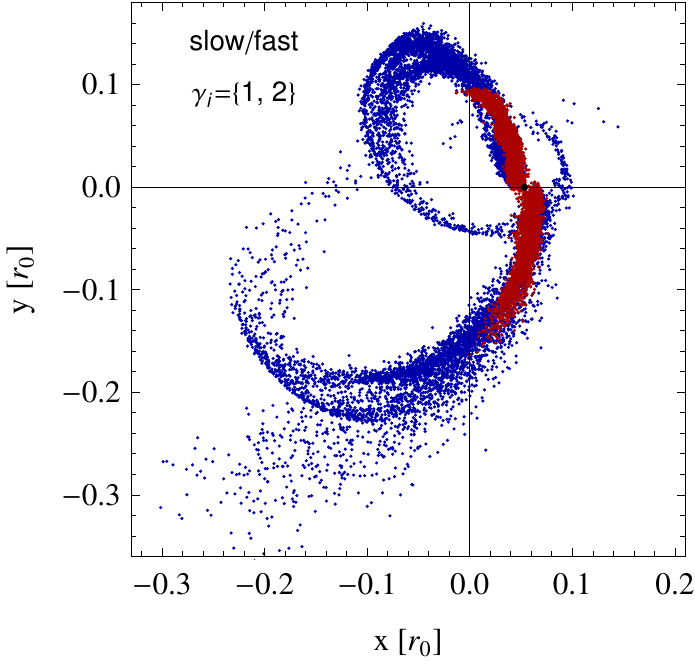}
\includegraphics[width=.347\textwidth]{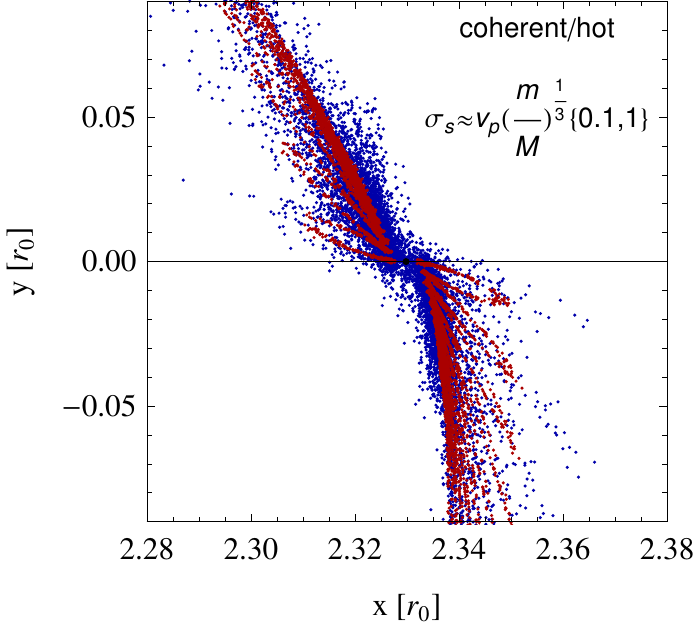}
\includegraphics[width=.314\textwidth]{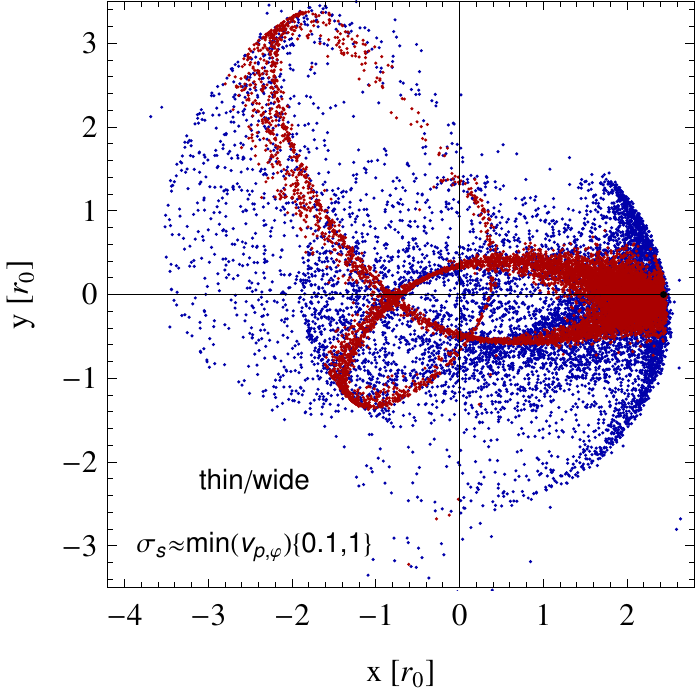}
\caption{An illustration of the three dichotomies that define the global properties of a tidal stream. In each panel two streams are superimposed with different 
colours to compare differences generated by the variation of a single physical quantity, all other properties being the same. Progenitors of the two streams 
overlap in all panels and are displayed as a black dot.
{\it Left panel}: after three pericentric passages, the length of two otherwise identical streams is dramatically different as a consequence of the
host's inner density slope $\gamma_i$ (see eqn.~(\ref{dkappaasym})), changing between  $\gamma_i=1$ in red and  $\gamma_i=2$ in blue (see Sect.~2.1).
{\it Central panel}: the magnitude of the spread in the escape velocities is varied, allowing the stream of a star cluster ($m/M\approx 10^{-8}$) to swap between a coherent and a hot regime, as prescribed by eqn.~(\ref{disrupt2}). The coherent stream in red, corresponding to conditions~(\ref{evap}) and~(\ref{evap2}) is essentially a one-dimensional streakline; internal coherence is lost in the opposite regime, when eqn.~(\ref{disrupt}) is satisfied (see Sect.~2.2).
{\it Right panel}: the magnitude of the spread in the escape velocities is varied with respect to the angular velocity of the progenitor at apocenter. As prescribed by eqn.~(\ref{shellsc2}), this ordering dominates the appearance of wide opening angles in the orbital plane (see Sect.~2.3).}
\label{regimex}
\end{figure*}

Three dichotomies between key physical quantities are identified and the mechanisms through which they regulate
the main qualitative properties of a stream are illustrated. Independently, they oppose:
\begin{itemize}
\item{tidal features in which differential streaming is slow (fast), i.e. that grow short (long) tails within one orbital period of the progenitor;}
\item{streams in which the internal dynamics is coherent enough to result in the formation of substructures, like `feathers' and `bifurcations', 
against streams in which members are internally well mixed;}
\item{streams that appear as such, i.e. thin and elongated within the orbital plane, against streams with large in-plane opening angles, or 
out-of-plane `umbrellas' and `shells'.}
\end{itemize}
{ These are of course only limiting cases (see Fig.~1), and each dichotomy is in fact associated with a dimensionless
ratio, illustrating a wide and continuous range of morphologies.}

Within this framework, I concentrate on the mechanism that causes the characteristic `feathering' of star cluster streams. 
The origin of such behaviour has been identified in the epicyclic motion of stars \citep{CD05, KA10, KA12}, which
is held responsible for the substructures and overdensities observed, for example, in the tidal tails of the GCs Palomar~5 (Pal~5) and GD1 
\citep[e.g.,][]{Ca12, Ca13}. Here, I show that, in the general case of non-circular orbits of the progenitor, the appearance 
of such `feathers' is in fact due to a time-modulation of the average mechanical energy of the escaping stars.
This modulation implies that particles released at different times stream away from each other with different
speeds, resulting in the formation of substructures and overdensities.
I investigate on the range of properties of host and progenitor that are necessary for such feathering to form. 
This is relevant to contextualise the multiple tidal tails observed around the ultrafaint dwarf galaxy Willman~I \citep{WB06},
the complex internal structure of the Anticenter stream \citep{Gr06c, Ca10} but also to understand { whether similar structures 
are to be expected around more ultrafaints}.

Another form of internal substructure is observed in the Sagittarius stream,  which displays a well defined bimodality in the stellar 
density of both leading and trailing arm \citep{BV06c, KS12}, a morphology that has so far eluded a satisfying explanation. 
Several mechanisms have been proposed as a solution to this puzzle, mainly belonging to two different families: the 
density peaks in the bimodal density pattern of each arm (i) originate from different progenitors \citep{KS12}, (ii) are 
misaligned tails shed at successive shedding events, corresponding to successive pericentric passages \citep{Fe06, PJ10}.
However, scenarios presented so far fall short of explaining the entirety of the observed phenomenology.

I show that the internal structure of each tidal arm, i.e. its streakline (see eqn.~(\ref{streakline}) for a definition), is naturally folded along most of its length. 
This folding is intrinsic, not produced by a shift of tails shed at different pericentric passages,
and is a result of the same energy modulation associated with the formation of feathers. This can potentially result in a 
bifurcation, i.e. in an evident bimodal density distribution, or be completely smeared out by the random motions 
internal to the stream itself, which are in turn an expression of the escape conditions. 
For a spherical Milky Way potential and a progenitor with no internal rotation, this mechanism can only produce
bifurcations that lie within the orbital plane. However, Sagittarius' bifurcation does not lie in the orbital plane.
Although the purpose of this paper is not to provide a model of Sagittarius, the conditions under which it would manifest 
an intrinsic bifurcation (as opposed to an artificial bifurcation caused by 
multiple wrappings) are explored.

In this paper, I use both analytical and numerical tools. In particular, I introduce a simple but very flexible model
for generating tidal streams that orbit within a spherically symmetric, { static} gravitational potential. This framework is similar to the 
`streakline method' used by \citet{KA12} and \citet{BA14}, and to the technique adopted by \citet{GS14}, in that, as in the mentioned works, 
particles are released along the progenitor's orbit. This model has been shown to reproduce the tracks and 
qualitative morphologies of both thin clusters' \citep[][]{KA12} and dwarfs' \citep[][]{GS14} streams. Using the insight gained
by the analysis of the physical mechanism responsible for the main morphologies of streams, I devise a model that can
describe the full phase space structure of a wider range of tidal debris. This is obtained by allowing for increased 
flexibility in the details of the escape conditions. We can then mimic: (i) the disruption of progenitors in a much wider range of mass ratios $m/M$ 
(where m and M are respectively the progenitors and host mass), by varying the probability function of the phase space conditions at escape; 
(ii) the disruption of an internally rotating progenitor, by modulation of the kick velocities; (iii) the details of both thin structured GCs' streams and
wide umbrellas and shells.

The present paper is structured as follows. 
In Sect.~2, I introduce the basic physical ingredients that shape the properties of a tidal streamer, and identify how their interplay 
define different dynamical regimes. In Sect.~3, I present the model for generating tidal features, and explore on the effects of the
progenitor's kinematics and shedding history. In Sect.~4, I concentrate on the dynamics of coherent streams, and illustrate the mechanism
that determines the formation of feathers and bifurcations. Sect.~5 presents qualitative applications to a few Milky Way streams. 
Sect.~6 lays the Conclusions of this work.

\section[]{Physical ingredients and dynamical regimes}
Tidal features are generated by the differential streaming of stars that have escaped the gravitational pull of their progenitor because of tidal 
forces. A statistical description of the escape process is not simple, as the phase space coordinates of stars at escape depend
on all the following: the mass profile of the host, the one of the progenitor, the progenitor's orbit as well as  
the star's orbit within the disrupting satellite before escape \citep[see e.g.,][and references therein]{RJ06, DE10}.
{ We can start by considering that escaping stars leave the progenitor }-- 
instantaneously at $(\bm{\rp},\bm{\vp})$ -- with a range of spatial and kinematical displacements:
\begin{equation}
(\bm{\rs},\bm{\vs})=(\bm{\rp},\bm{\vp})+(\bm{\delta r},\bm{\delta v})\ .
\label{displ}
\end{equation}
The subsequent evolution of the escapees is determined by the Hamiltonian flow of the combined gravitational potential of host, progenitor
and escaping material. Such flow propagates the initial phase space displacements of the tidally stripped stars, a process that appears 
as a differential streaming in physical space, and that progressively mixes the corse-grained distribution function in phase space \citep{HA99}.

{ A closer description of the conditions that allow stars to leave their progenitor has been the subject of significant attention in the literature, 
 for example in relation to the evolution of star clusters} \citep[see e.g.][and references therein]{TF00,RF11}.
{ Shed stars escape the progenitor where its gravitational attraction is balanced by the tidal forces. This condition is dependent on the value of the
Jacobi `constant' $E_{\rm J}$ of each star\footnote{The term `constant' nominally applies to circular progenitors obits only, but it is still useful for a qualitative description.}. 
Only stars with 
\begin{equation}
E_{\rm J}\geq \bar{E}_{\rm J}\ 
\label{jacobi}
\end{equation}
are allowed to infinite distances from their progenitor. Stars that have just enough energy to escape, $E_{\rm J}= \bar{E}_{\rm J}$,
are forced to do so through the only apertures of the 
Jacobi surface, the so called Lagrange points, which define the instantaneous tidal radius
\begin{equation}
\rt = \left({{G m}\over{\Omega^2-{\partial}^2 \Phi /{\partial}r^2}}\right)^{1\over3}\ ,
\label{rt}
\end{equation}
with $\Phi$ being the host's gravitational potential and $\Omega$ its angular frequency.
The size of these apertures around the Lagrange points increase with the value of the Jacobi constant, 
so that more energetic stars escape more easily, and with a wider range of initial conditions}
\begin{equation}
\bm{\delta r}\equiv \st\ \rt\ \hat{\bm r}_p + \bm{\varpi}\ ,
\label{rprime}
\end{equation}
{ where $\st\in \left\{ -1,+1 \right\}$ identifies the leading and trailing condition.
If $ \bm{\varpi}=0$, stars escape exactly from the instantaneous tidal radius. }

The tidal radius scales like $(m/M)^{1/3}$, where $m$ is the mass of the satellite and $M$ is the host's mass and, 
as recognised by \citet{JK98}, it sets a natural energy scale of the shedding mechanism:
\begin{equation}
\delta E_{\rm t} \equiv \Phi(\bm{\rs})-\Phi(\bm{\rp}) \approx \st\ \rt {{{\partial}\Phi}\over{{\partial}r}}\biggr|_{\bm{\rp}}\ ,
\label{dEt}
\end{equation}
this is the (instantaneous) difference in gravitational energy between the progenitor and a star that 
escapes from the vicinity of either saddle points of the effective potential. Additionally, the escapee's shift in velocity gives rise to a companion 
shift in mechanical energy:
\begin{equation}
2\ \delta E_{\rm k} \equiv ||\bm{\vp} + \bm{\delta v}||^2 - ||\bm{\vp}||^2 \approx 2\ \bm{\vp}\cdot \bm{\delta v} \ ,
\label{dEk}
\end{equation}
so that the total energetic difference between progenitor and escaped star is 
\begin{equation}
\delta E = \delta E_{\rm t} + \delta E_{\rm k}\ .
\label{dE}
\end{equation}

Although the dynamical dichotomies I set out in the following are valid in the general case, for simplicity of 
description and to favour a more explicit identification of the involved physical quantities, I assume that the host's gravitational potential is spherical. 
This implies that orbits are uniquely identified by their energy and angular momentum. In the following, I use alternatively $(E, J)$ or $(E, j)$ pairs, 
where $0\leq j \leq1$ is the usual circularity:
\begin{equation}
j = j(E,J)\equiv J/J_{\rm c}(E) \ 
\label{circ}
\end{equation}
and $J_{\rm c}(E)$ is of course the angular momentum of the circular orbit with energy $E$. 
Also, { while the progenitor clearly influences the long-term evolution of the tidal stream by shaping the distribution of the phase 
space displacements $(\bm{\delta r},\bm{\delta v})$, I discard its gravitational influence once debris are unbound,
and the mutual gravitational influence of other debris}. 

In this simplified framework it is easy to right down the angular momenta of progenitor ${\bm \Jp}$ and escaped star ${\bm \Js}$.
{ At first order}
\begin{equation}
{\bm \Js}\approx{\bm \Jp}\left(1+\st{\rt\over\rp}\right)+ {\bm \vp}\wedge{\bm \varpi}+{\bm \delta v}\wedge{\bm \rp} \ ,
\label{deltaJ}
\end{equation}
{ so that ${\bm \Jp}$ and ${\bm \Js}$ differ in both modulus and direction}.
%
%
{ Changes in modulus $\Js$ mainly affect orbital shape in the host's potential,}
{ while changes in the direction $\hat{\bm J}_{\rm s}$ tilt the orbital plane of each shed star.
More quantitatively, the escapee's orbital plane is rotated with respect to the progenitor's one by an angle}
\begin{equation}
\omega=\arccos\left({{\hat{\bm J}_{\rm s}\cdot \hat{\bm J}_{\rm p}}}\right)\ ,
\label{oplanetilt}
\end{equation}
{ around the direction of the line of nodes}
\begin{equation}
{\bm {n}}={{{{\bm {\Js}}\wedge {\bm {\Jp}}}}}\ .
\label{nodes}
\end{equation}
\begin{figure}
\centering
\includegraphics[width=.8\columnwidth]{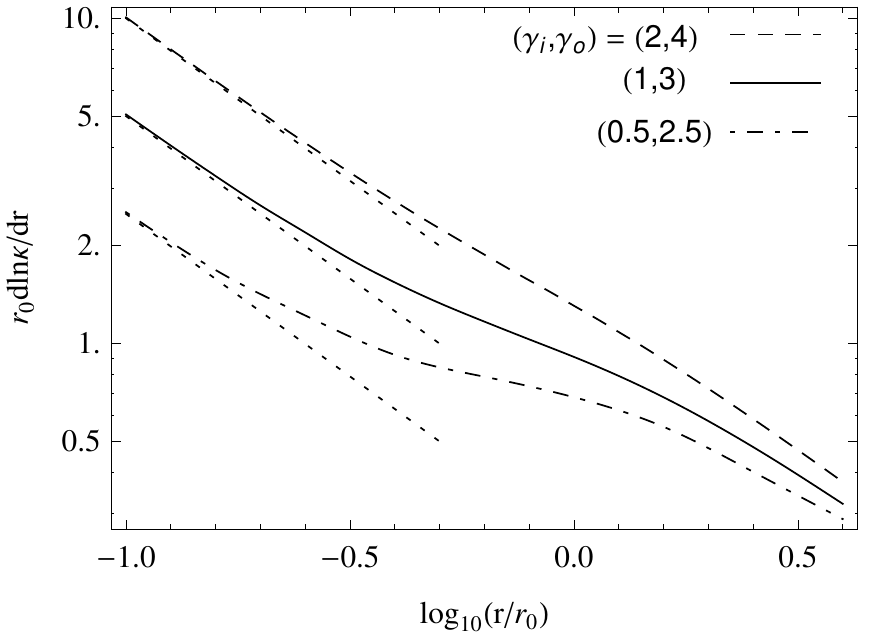}
\caption{The quantity $r_0{\rm d ln}\kappa/dr$ for gravitational potentials generated by broken power-law density profiles, eqn.~(\ref{bpwl}). This is proportional to the 
(relative) angular growth speed of the stream, as in eqn.~(\ref{kappaord}). Therefore, centrally steeper host density profiles and smaller pericentric radii allow for a faster development of tidal features (see Sect.~2.1). Dotted lines provide asymptotic approximations for the central pure power-law regime.}
\label{dkappa}
\end{figure}

\subsection{Slow vs fast angular mixing}
The first dichotomy I am going to consider is defined by the comparison between the quantities
\begin{itemize}
\item{$\langle\Omega_{\rm p}\rangle=\Delta \varphi_{\rm p}/T_{\rm r,p}$, i.e. the average orbital angular frequency of the disrupting progenitor.
Here, $T_{\rm r,p}=T_{\rm r}(\Ep,\jp)$ is the progenitor's radial period and $\Delta \varphi_{\rm p}$ is the angle the progenitor spans in such time.}
\item{$\langle \delta \Omega\rangle=\langle\Omega_{\rm s}\rangle-\langle\Omega_{\rm p}\rangle$, i.e. the average difference in orbital frequency 
of shed material and progenitor. Here, $\langle\Omega_{\rm s}\rangle$ is the average angular frequency
of a given collection of shed particles, for example those released at pericenter and belonging to the leading (or trailing) tail.}
\end{itemize}

Streams for which 
\begin{equation}
\langle \delta \Omega\rangle \ll \langle\Omega_{\rm p}\rangle
\label{slowc}
\end{equation}
are streams in which differential streaming is slow and many orbits of the progenitor are required before long tails can form.
On the other hand, the tails of streams for which 
\begin{equation}
\langle \delta \Omega\rangle \gtrsim \langle\Omega_{\rm p}\rangle
\label{fastc}
\end{equation}
complete one or more wraps for each progenitor's orbital time and are characterised by a much faster evolution.

The differential angle between an escaped star and its progenitor, after a time $t$ since escape, has the secular component (i.e. ignoring the details of each orbital oscillation)
\begin{equation}
\varphi_{\rm s}(t)-\varphi_{\rm p}(t)\approx t\left({{\Delta \varphi_{\rm s}}\over{T_{\rm r,s}}}-{{\Delta \varphi_{\rm p}}\over{T_{\rm r,p}}}\right)\approx t\ \bm{\nabla}\left({{\Delta \varphi}\over{T_{\rm r}}}\right)\cdot \binom{\delta E}{\delta j}
\label{phid}
\end{equation}
where the gradient $\bm\nabla$ refers to derivatives with respect to energy and circularity: $\bm\nabla_{(E,j)}$.
As recognised by \citet{JK98}, the azimuthal angle $\Delta \varphi$ is essentially a function of the angular momentum only, 
while the radial period is more strongly dependent on the orbital energy, so that eqn.~(\ref{phid}) can be well approximated by
\begin{equation}
\varphi_{\rm s}(t)-\varphi_{\rm p}(t)\approx t\ \left[{1\over{T_{\rm r,p}}} {{\partial \Delta \varphi}\over{\partial j}} \delta j + \Delta \varphi_{\rm p} {{\partial (1/T_{\rm r})} \over{\partial E}}\delta E\right]\ .
\label{phid2}
\end{equation}
In normal conditions, the energy term is significantly larger than the one arising from differences in the circularity, which implies that
particles shed at similar times end up ordered in the stream according to their orbital energy \citep[e.g.,][and the following Sect.~4]{JK01}.
As a consequence, we have that
\begin{equation}
{{\langle\delta \Omega\rangle}\over{\langle\Omega_{\rm p}\rangle}}\approx{\rt {T_{\rm r,p}}}{{\partial}\over{\partial r}}{1\over T_{\rm r}}\bigg{|}_{\rp}  \approx{1\over \kappa} {{{\rm d}\kappa}\over{{\rm d}r}}\bigg{|}_{\rp}\rt\ ,
\label{kappaord}
\end{equation}
in which $\kappa$ is the usual epicyclic frequency and I have used that $\delta E(\partial r/\partial E )\approx \rt$, and that $T_{\rm r}(E,j)\approx T_{\rm r}(E,1)$. 

Assuming for example that the host's mass within pericenter and the progenitor's mass are both fixed -- hence $\rt$ is approximately fixed -- eqn.~(\ref{kappaord})
shows that the speed of the differential streaming is dominated by the slope of the host's gravitational potential \citep[][explore a similar issue, 
but within the complementary framework of the action-angle formalism]{BE14}. Fig.~\ref{dkappa} shows the magnitude of the function $r_0/\kappa\cdot {\rm d}\kappa/{\rm d}r$ 
for a few gravitational potentials, generated from density profiles within the family of broken power-law models:
\begin{equation}
\rho(r)=\rho_0 \left({r\over r_0}\right)^{-\gamma_i}\left[1+\left({r\over r_0}\right)^2\right]^{-{{(\gamma_o-\gamma_i)}/ 2}}\ .
\label{bpwl}
\end{equation}
Especially in the central regions, steeper density profiles determine substantially faster mixing times.
This can be easily understood by considering the asymptotic expansion in the pure power-law regime $r/r_0\ll 1$:
\begin{equation}
{1\over \kappa} {{{\rm d}\kappa}\over{{\rm d}r}}\sim {\gamma_i\over 2}{1\over r}\ ,
\label{dkappaasy}
\end{equation}
displayed in Fig.~\ref{dkappa} as dotted lines. Therefore, we see that, for a given $\rt$, the speed of differential streaming
and consequent phase space mixing is directly proportional to the host's inner density slope $\gamma_i$. 

The left panel of Fig.~(\ref{regimex}) illustrates the validity of eqn.~(\ref{dkappaasy}) in a practical case. It shows two streams that are identical
except for the host's density slope, which varies between $\gamma_i=1$ and $\gamma_i=2$. The entire stream lives at $r\ll r_0$, so that 
the density profile is essentially a pure power law. Both snapshots are taken after three pericentric passages and three associated 
shedding events. However, it is evident that the two streams span substantially different angles. 
The strong and direct dependence of eqn.~(\ref{dkappaasy}) on the density slope
is extremely promising for future studies, especially if combined to a clear detection of a thin stream's feathering, 
which instead provides a measurement of the time passed since escape (see Sect.~4). 

Finally, it can be noted that in the limit of a cored density profile, $\gamma_i=0$, phase mixing is substantially slowed down by the
characteristic solid-body behaviour of the relevant frequencies, which inhibits any differential streaming. Tracer particles slosh 
back and forth within the harmonic region of the potential without any substantial mixing. This provides an 
analytical interpretation to the results of \citet{Kl03, SS10, LV13}. The survival times of cold kinematic clumps 
in dwarf galaxies are systematically longer when the halo density profile is cored.

On the other hand, by assuming that the host's mass within the orbital pericenter $M$ is fixed,
and using that eqn.~(\ref{rt}) implies the scaling
\begin{equation}
\rt\sim \rp\ \left({m\over {\gamma_i M}}\right)^{1\over3}\ ,
\label{rt2}
\end{equation}
we can highlight the progenitor's mass scale associated with different angular mixing
regimes:
\begin{equation}
{{\langle\delta \Omega\rangle}\over{\langle\Omega_{\rm p}\rangle}}\approx{1\over 2}{\gamma_i^{2\over 3}}\left({m\over M}\right)^{1\over 3}\ .
\label{dkappaasym}
\end{equation}
GCs of the Milky Way will inevitably require tens of orbital times for their tails to extend for a fraction of their orbital azimuthal angles,
while the tails of dwarf satellites with $m/M\gtrsim10^{-2}$ can potentially wrap the Galaxy within one orbital time.

\subsection{Internal coherence vs hot streaming}
The physical regimes analysed in this Section deal with the coherence in the internal dynamics of the stream,
and can be separated out by comparing the magnitudes of the following two quantities:
\begin{itemize}
\item{$\langle \delta \Omega\rangle$, i.e. the previously defined average difference in orbital frequency of progenitor and material shed at some given time;}
\item{the spread in its distribution, within the same ensemble of shed material: }
\end{itemize}
\begin{equation}
\sigma(\delta \Omega)=(\langle\delta\Omega^2\rangle-\langle\delta\Omega \rangle^2)^{1/2} \ .
\label{sdomega}
\end{equation}
Streams for which 
\begin{equation}
\sigma(\delta\Omega)\ll\langle\delta\Omega\rangle
\label{evap}
\end{equation}
have tails in which the ordering defined by the shedding time is preserved by the Hamiltonian flow, i.e. 
mixing along the stream of particles released at different times is limited, allowing for the formation and 
survival of internal substructure.
On the other hand, if
\begin{equation}
\sigma(\delta\Omega)\gtrsim\langle\delta\Omega\rangle\ ,
\label{disrupt}
\end{equation}
the ordering of particles is dominated by mechanical energy (rather than release time),
and tails appear warmer. 

Note that, the limit of eqn.~({\ref{evap}}) is the same limit that defines a `streakline' (see eqn.~(\ref{streakline}) for a definition). 
In fact, if the phase space displacements $(\bm{\delta r},\bm{\delta v})$ have no scatter 
(equivalently, the orbital pairs $(E_{\rm s},J_{\rm s})$ of the escapees have no scatter), they
define a one-dimensional manifold of initial conditions in phase space, whatever their dependence on time. 
The Hamiltonian flow preserves this one-dimensionality, so that such a stream and its streakline are exactly the same. 

As we have seen in Sect.~2.1, $\delta \Omega$ is essentially a function of the energy difference $\delta E$. Henceforth, the ordering
of this Section is equivalent to an ordering between $\langle\delta E\rangle$ and $\sigma(\delta E)$. In turn, the instantaneous 
spread $\sigma(\delta E)=\sigma(\delta E_{\rm t})+\sigma(\delta E_{\rm k})$ is a result of the distribution of the phase space 
coordinates of stars at escape $(\bm{\delta r},\bm{\delta v})$, or $(\bm{\varpi},\bm{\delta v})$ as far as spreads are concerned.
The necessary condition for coherence can be safely derived by discarding the energy spread 
resulting from the distribution in the `spatial' escape conditions, i.e. by assuming for the moment $\sigma(\delta E_{\rm t})=0$.
By order of magnitude, random velocities at escape imply the energy spread
\begin{equation}
\sigma(\delta E_{\rm k})\approx \bm{\vp}\cdot\sigma(\bm{\delta v})\approx \vp\  \sigma_{\rm s}\ , 
\label{kinEspread}
\end{equation}
where I have defined $\sigma_{\rm s}$ as the characteristic velocity spread of the escaping stars. 
On the other hand,
\begin{equation}
\langle\delta E\rangle \approx \delta E_{\rm t}\approx \vp^2  \left({m\over M}\right)^{1\over3}
\label{kinEspread2}
\end{equation}
so that conditions~(\ref{evap}) (and~(\ref{disrupt})) are respectively equivalent to
\begin{equation}
\sigma_{\rm s}\ll\ (\gtrsim)\ \vp \left({m\over M}\right)^{1\over3}\ .
\label{disrupt2}
\end{equation}
These inequalities set the divide between ordered, essentially one-dimensional streams that closely follow their streaklines and 
streams in which the internal dynamics is dominated by a substantial energy-driven differential streaming.

The central panel of Fig.~\ref{regimex} shows the streams generated by a progenitor having $m/M\approx10^{-8}$ 
(at pericenter), after it has been shedding stars for over ten orbital times. 
All parameters are kept fixed between the two realisations, apart for the velocity dispersion at escape $\sigma_{\rm s}$.
As found in eqn.~(\ref{disrupt2}), the magnitude of the orbital velocity, weighted by the progenitor-to-host mass ratio,
drives the transition between the two opposed regimes just described. The internally cold stream essentially behaves like its 
one-dimensional streakline and manifests the typical feathering, while this coherence is destroyed by random motions in the 
second stream (in blue).

As mentioned earlier, condition~(\ref{disrupt2}) remains valid when the contribution of $\sigma(\delta E_{\rm t})$ is not discarded.
In order to show this, I need to take into account that the phase space coordinates at escape $\delta v$ and $ \varpi$
are correlated (see Section~2). I can estimate the size and shape of these apertures by considering the behaviour of the Hills surfaces
\begin{equation}
\Phi_{eff}(\bm{r})=\bar{E}_{\rm J}+{|\bm{\delta v}|^2\over2}
\label{semiax2}
\end{equation} 
in the vicinity of the tidal radius, as a function of $\delta v$. Here, $\Phi_{eff}$ is the `efficient' potential experienced by each 
particle in the reference frame that moves together with the progenitor. To make the problem analytically tractable, I assume that
the progenitor's orbit is circular, the host potential is a scale free power law ($r\ll r_0$), and that $|\bm{\delta v}|^2\ll|\bar{E}_{\rm J}|$, 
or equivalently that $ \varpi\ll \rt$. In this a case, a particle that reaches the tidal radius with a velocity $\delta v$ can escape from 
an elliptical aperture (defined in the plane crossing $\rt$ and perpendicular to the direction $\bm\hat r$)
with semiaxes -- respectively in the directions $\bm{\hat\varphi}$ in the orbital plane and $\bm{\hat z}$, perpendicular to the orbital plane,
\begin{align}
\bar\varpi_{\varphi} & =  \rt \sqrt{{\rt |\bm{\delta v}|^2}\over{G m}}\ , \label{semiax1}\\
\bar\varpi_{z} & = \bar{\varpi}_{\varphi}   \sqrt{\gamma_i \over{\gamma_i+1}}\ . \label{semiax2}
\end{align} 
These assume that the gravitational field of the progenitor is Keplerian, although a different description would only 
affect coefficients of order unity rather than change dimensional dependences. As the resulting relative energetic spread is
\begin{equation}
{{\sigma (\delta E_{\rm t}) }\over{\langle\delta E_{\rm t}\rangle}}\approx {{\sigma(\varpi)}\over\rt} \ ,
\end{equation}
it is easy to see that the scalings of eqn.~(\ref{semiax1}) and~(\ref{semiax2}) imply that 
\begin{equation}
{{\sigma (\delta E_{\rm t}) }\over{\langle \delta E_{\rm t}\rangle}}\approx {{\sigma (\delta E_{\rm k}) }\over{\langle \delta E_{\rm k}\rangle}} \sim{\sigma_{\rm s}\over{v_{\rm p}(m/M)^{1/3}}}\ ,
\label{disrupt3}
\end{equation}
proving that  the simplification used earlier to derive eqn.~(\ref{disrupt2}) was indeed safe.

%
%

It should be noted that, differently from the stream's length (see eqn.~(\ref{dkappaasym})) and despite the appearance of
eqn.~(\ref{disrupt2}), the mass ratio $m/M$ is nominally not a direct player here. Intuitively, this is because more massive progenitors 
loose `warmer' particles due to their higher internal velocity dispersions, but also have faster streams, due to their larger tidal radii.
In fact, it is a simple exercise to see that all dimensional dependences in eqn.~(\ref{disrupt2}) simplify once they are made explicit, 
for example by using the approximation
\begin{equation}
\sigma_{\rm s}\approx \left({{mG}\over{k \rt}}\right)^{1\over2}\ ,
\label{sigmavs}
\end{equation}
where $k$ is a fudge factor similar to the well known virial coefficient. $k$ usually stands for the structural properties of a gravitating structure, 
but, in this occasion, its meaning is crucially enriched by the relative size of progenitor and tidal radius $r_{\rm h}/\rt$ -- we use $r_{\rm h}$ to
indicate the half light radius of the progenitor. 
%
%

{ In particular, if a progenitor is sufficiently extended, the nominal tidal radius 
$\rt$ as in eqn.~(\ref{rt}) looses its meaning, as $\rt$ goes to zero as $\rp$ goes to zero, but the escape regions do not.
When
 \begin{equation}
\rt\lesssim r_{\rm h} \ ,
 \label{hot0}
 \end{equation}
particles are lost from wide range of escape conditions, 
 \begin{equation}
\varpi\approx r_{\rm h} \ ,
 \label{hot1}
 \end{equation}
resulting in the progenitor disrupting with a considerable $\sigma(\delta E)$.}
In conclusion, the progenitor's mass is not a direct player in determining the internal coherence of a stream only to the point
the progenitor is more compact than its nominal tidal radius. However, as mass and characteristic 
size are certainly correlated, the systematic influence of mass is played through the dimensionless ratio $r_{\rm h}/\rt$.


We can then envision two extremes.
On one hand there are the so called `evaporative conditions' \citep[see also][]{KA12}, while the opposite case is represented by an 
almost impulsive, hot disruption. In the evaporative case, shed stars are essentially `peeled off' \citep{BJ14}: they escape with 
\begin{equation}
{E_{\rm J} - \bar{E}_{\rm J}\over{| \bar{E}_{\rm J}|}}\ll 1
\label{evap2}
\end{equation}
or equivalently satisfying eqn.~(\ref{disrupt2}), 
so that they have very little spread in the phase space displacement, both spatial and kinematical. 
This is for example the case of a stellar cluster that 
does not fill its Roche lobe. Rather than directly because of tides, stars gain enough energy to escape due to internal 
collisional evolution: the escape rate is almost constant, with little modulation on the orbital period. On the other hand, 
\begin{equation}
{E_{\rm J} - \bar{E}_{\rm J}\over{|\bar{E}_{\rm J}}|}\gtrsim 1
\label{hot2}
\end{equation}
results in warmer streams: stars escape from wider regions, and with a variety of initial kinematic displacements, 
contributing to a large spread in both $\delta E_{\rm t}$ and $\delta E_{\rm k}$. This is for example the case of a system 
on an eccentric orbit, and diffuse enough to satisfy condition~(\ref{hot0}). 
Shedding is likely not constant in time, with strong modulation with the orbital period and considerable mass loss around pericenter. 

\begin{figure}
\centering
\includegraphics[width=.8\columnwidth]{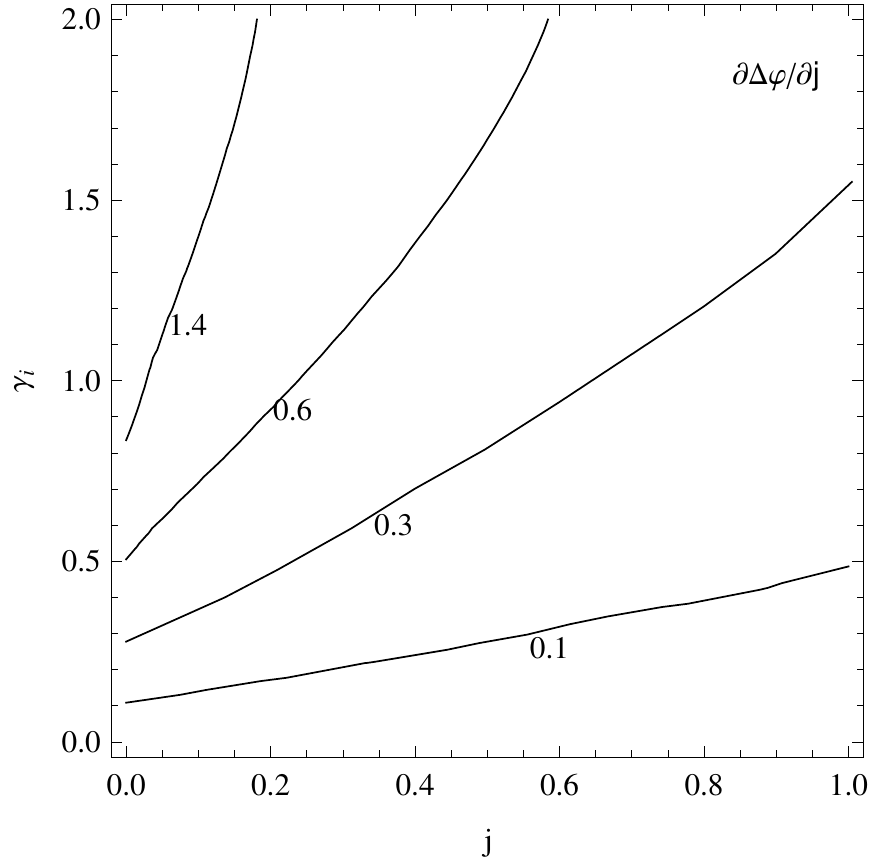}
\caption{The contours of the quantity $\partial\Delta\varphi/\partial j$ in the pure power law regime, for varying
power law indexes $\gamma_i$ and circularities $j$. This is proportional to the 
angular size of the stream in the orbital plane, as in eqn.~(\ref{shellang}). Therefore, centrally steeper host density profiles and 
more eccentric progenitor's orbits allow for the formation of wider streams (see Sect.~2.3).}
\label{dphidj}
\end{figure}

\subsection{Thin streams vs wide streams and shells}

I advance in this synopsis towards ever `warmer' tidal structures by considering a third and last dichotomy. This is 
based on the term of eqn.~(\ref{phid2}) that depends on the variations in the circularity of the shed material, and that I 
have previously neglected in the Sect.~(2.1), as subdominant with respect to the companion energy term. This subdominance assures 
that an ensemble of particles that have been shed with approximately the same energy will experience a similar differential 
streaming with respect to the progenitor, so to define a single position along the tail. 
However, with respect to one another, they satisfy $\delta E=0$ and the way these particles 
stream away from each other is in fact determined by the circularity term. Therefore, by comparing
\begin{itemize}
\item{the spread in the term $\delta j({{\partial \Delta \varphi}/{\partial j}})$ over an ensemble of particles shed with the same energy,}
\item{with unity,}
\end{itemize}
I am comparing streams of different widths in the orbital plane. 
Streams satisfying the conditions
\begin{equation}
\sigma(\delta j\big{|}_E){{\partial \Delta \varphi}\over{\partial j}}\biggr{|}_{(\Ep,\jp)}\ll1
\label{thinc}
\end{equation}
appear thin, while the opposite case
\begin{equation}
\sigma(\delta j\big{|}_E){{\partial \Delta \varphi}\over{\partial j}}\biggr{|}_{(\Ep,\jp)}\gtrsim1
\label{widec}
\end{equation}
is characterised by substantial opening angles. In particular, using eqn.~(\ref{phid2}), I get that the spread above 
provides an estimate for the angle covered by particles at their first trailing/leading apocenters, and is then an approximate
measure of the opening angle of a stream within the orbital plane:
\begin{equation}
\sigma(\delta j\big{|}_E){{\partial \Delta \varphi}\over{\partial j}}\biggr{|}_{(\Ep,\jp)}\approx \sigma\left[\varphi_{\rm s}(T_{\rm r,p})\right]\ .
\label{shellang}
\end{equation}

Fig.~\ref{dphidj} shows the influence on the current ordering provided by the properties of the host potential. 
For different circularities, it displays the contours of the quantity ${{\partial \Delta \varphi}/{\partial j}}$ for gravitational potentials generated by the 
density profiles of eqn.~(\ref{bpwl}), in the scale-free pure power law regime. Clearly, for a fixed spread $\sigma(\delta j\big{|}_E)$, progenitor 
orbits that are more strongly radial tend to generate streams that are wider in the orbital plane. Also, steeper host density profiles
have an analogous effect. 
On the opposite side, 
as the harmonic limit has $\Delta \varphi(j)=\pi$ for all circularities, imposing $\gamma_i=0$ also prevents angular momentum driven
differential streaming. 

%
%
I first concentrate on progenitors that do not get too close to the centre of their host, $\rh\ll \rp$, so that 
$\delta J\approx\delta J_{\rm k}\approx \rp \delta v_{\varphi}$. It follows that the width of the stream in the orbital plane is primarily an expression of
random motions escape, through $\delta v_{\varphi}$. Wide opening angles are obtained for $\sigma(\delta j\big{|}_E)\approx 1$, or equivalently for
\begin{equation}
\sigma_{\rm s,\varphi}\gtrsim v_{\rm p,\varphi}\ .
\label{shellsc2}
\end{equation}
This provides an interesting comparison with eqn.~(\ref{disrupt2}): wide in plane streams are formed after internal coherence is lost, 
for even larger values of the spread in the kick velocities. The right panel of Fig.~\ref{regimex} shows two streams, which are exactly the same except 
for the velocity dispersion $\sigma_{\rm s,\varphi}$, evolved for two pericentric passages. As the random motion of stars at escape
become comparable in magnitude to the orbital velocity of the progenitor at apocenter, the opening angles in the orbital plane grow accordingly.
Also note that, as the orbital velocity $v_{\rm p,\varphi}$ is periodic, material that 
is shed along eccentric orbits is more likely to comply with eqn.~(\ref{shellsc2}) when it escapes near apocenter.

In the direction perpendicular to the orbital plane, $\bm{\hat z}$, the width of the stream is instead determined by the distribution 
of orbital planes of its members, i.e. by the distribution of the directions of the angular momenta ${\bm\Js}$,
as prescribed by eqns.~(\ref{deltaJ}),~(\ref{oplanetilt}) and~(\ref{nodes}).
Therefore, the angular width of a stream in the direction $\bm{\hat z}$ can be connected to the
dispersion $\sigma_{{\rm s}, z}$:
\begin{equation}
\sigma(\omega)\approx \tan \left( \sigma_{{\rm s}, z}\over v_{{\rm p}, \varphi}\right).
\label{widthz}
\end{equation}

\begin{figure}
\centering
\includegraphics[width=.52\columnwidth]{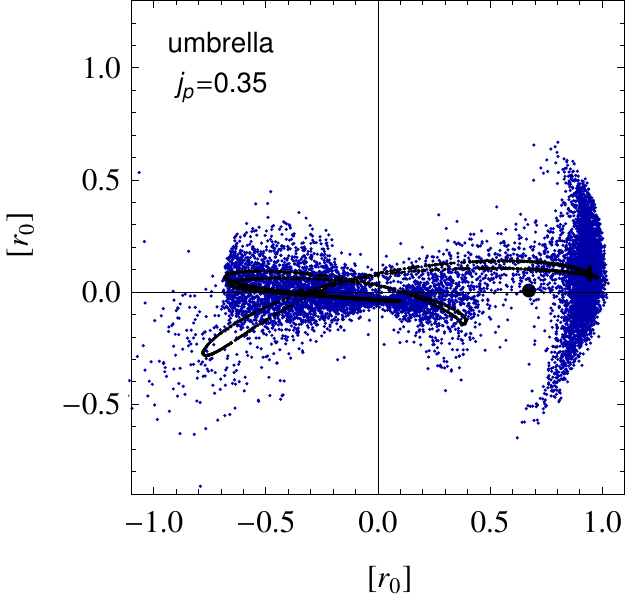}\hspace{-.15in}
\includegraphics[width=.514\columnwidth]{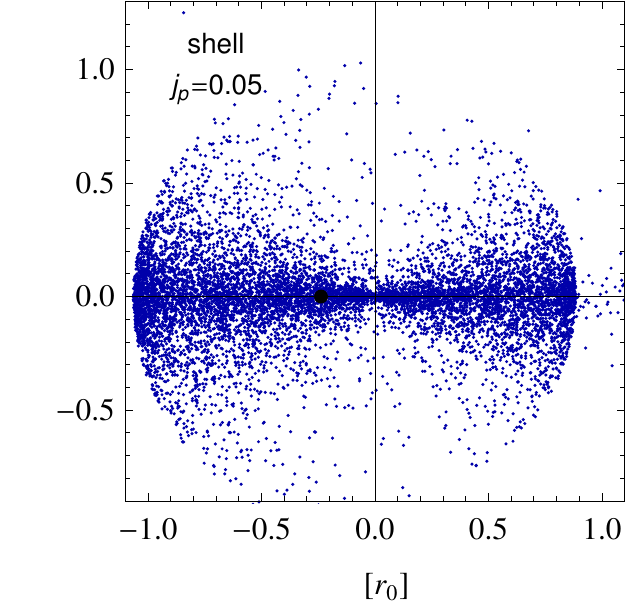}
\caption{{\it Left:} an example of an `umbrella', obtained at $\jp=0.35$, with a small enough pericenter to 
result in wide out-of-plane opening angles, although with a substantial orbital angular momentum; the plane 
of progenitor's orbit and streakline -- shown in black -- is almost edge on. 
{\it Right:} an example of a `shell', obtained for an almost exactly radial orbit $\jp=0.05$; the streakline is not visible 
as the orbital plane is exactly edge on.   }
\label{ombshe}
\end{figure}

These scalings loose their validity when the progenitor's orbit is exceedingly eccentric, and
\begin{equation}
\rp\lesssim \rh \ ,
\label{shellsc3}
\end{equation}
implying that 
\begin{equation}
\varpi\approx \rp \ .
\label{shellsc0}
\end{equation}
In this case, { eqn.~(\ref{deltaJ}) shows that, through $\bm{\delta J_{\rm t}}$, particles are lost with a significant scatter 
in the direction of the angular momentum vector ${\bm \Js}$.
This is mirrored in a variety of orientations for the orbital planes of the escapees, so that the opening angle of the tidal debris in 
the direction perpendicular to the orbital plane is sizeable: $\sigma(\omega)\gtrsim1$. 
As illustrated in Fig.~\ref{ombshe}, this is the regime of wide umbrellas and concentric shells.}

\section[]{streams in spherical potentials: a fast and flexible approach }

As mentioned in the Introduction, the idea at the basis of the stream generating model I present here is  
similar to the one used by \citet{KA12} and \citet{GS14}: individual particles are released along the 
progenitor's orbit, and their distribution in phase space at a later time samples the properties of the stellar stream.
{ This technique is usually referred to as the `streakline method'. If $H$ is the Hamiltonian flow and $({\bm{x_0}},{\bm{v_0}}, t_0)$ 
are initial conditions, a streakline is the locus}
\begin{equation}
\mathcal{S}=H(t; {\bm{x_0}},{\bm{v_0}}, t_0)\  \ {\rm for}\  \  t_0\in[t_1, t_2]\ ,
\label{streakline}
\end{equation}
{ i.e. the ensemble of all particles that have departed a given $({\bm{x_0}},{\bm{v_0}})$ between time $t_1$ and $t_2$, evolved to time $t$.
In the case of tidal debris, rather than a fixed point $({\bm{x_0}},{\bm{v_0}})$ in phase space, initial conditions are associated with the instantaneous position and velocity
of the (leading or trailing) tidal radius, from which particles abandon their progenitor.}

The main differences between previous implementations and the one I present here are that I can freely vary both
\begin{itemize}
\item{the modulation with the orbital time of the probability distribution of the shedding times;}
\item{the probability distribution of the phase space displacements at escape $({\bm {\delta r}},{\bm {\delta v}})$. }
\end{itemize}
This allows me to reproduce the phase space properties of streams across the entire range of 
physical regimes identified in Sect.~2. Furthermore, this generative method does not require solving the equations
of motion for each single stream member: escapees are evolved using orbit libraries as described in the following.

\subsection{Orbit libraries}

To parametrise the host's potential well, the present algorithm makes use of the family of broken power law density profiles~(\ref{bpwl}).
Inner and outer density slopes, $\gamma_i$ and $\gamma_o$, are allowed to independently vary within the intervals
\begin{equation}
\gamma_i\in[0,2.9]\  \  ;\  \  \gamma_o\in[2.1,5]\ ,
\label{slopes}
\end{equation}
which allows me to cover a wide variety of density profiles, useful when modelling streams in different contexts.
Within this family of gravitational potentials, the properties of orbits are conveniently stored in purposely 
optimised libraries. 

As I am restricting the analysis to spherical potentials, all functional dependences that 
are associated with time can be isolated to the $[0,1]$ interval. These span variations between the orbital 
apo- and peri-center. For example, if $(t,\varphi)=(0,0)$ indicates pericenter, both time and angular phase
can be replaced by the following rescaling
\begin{equation}
\bar\varphi=\biggr{|}{\varphi\over{\Delta\varphi/2}}-2\bigg\lfloor{{{\varphi-\Delta\varphi/2}\over{\Delta\varphi}}}\bigg\rfloor-2\biggr{|}\ ,
\label{aphase}
\end{equation}
which maps $\varphi$ in $[0,1]$ and where the symbol $\lfloor{\cdot}\rfloor$ indicates the usual floor function. The same
transformation can be used to normalise time, using the orbital period $T_{\rm r}$ in place of the angle $\Delta\varphi$
\begin{equation}
\bar t=\biggr{|}{t\over{T_r /2}}-2\bigg\lfloor{{{t-T_r /2}\over{T_r }}}\bigg\rfloor-2\biggr{|}\ ,
\label{atime}
\end{equation}

while it is possible to rescale the advancement of galactocentric distance between pericenter and apocenter using the following
\begin{equation}
\bar r={{r-r_{peri}}\over{r_{apo}-r_{peri}}}\ .
\label{aradius}
\end{equation}
In such a way, the core functions necessary to the orbit library
\begin{equation}
\{\bar{t}(\bar{r})\ ;\ \bar{r}(\bar{\varphi})\ ;\ \bar{\varphi}(\bar{t})\}
\label{dimlessfs}
\end{equation}
are functions that map $[0,1]\rightarrow[0,1]$, which makes numerical interpolation considerably more efficient.
Of course, all mentioned functions are also dependent on the host's potential -- through the density slopes $(\gamma_i,\gamma_o)$ -- 
and on the orbital pair $(E,j)$, for a total of five free variables. On the other hand, it is necessary to 
store the quantities that define the scalings above: apocentric and pericentric 
distances $(r_{apo}, r_{peri})$, azimuthal angles $\Delta\varphi$, radial periods $T_{\rm r}$. 
Though not of time, these are all functions of four variables: the density slopes $(\gamma_i,\gamma_o)$
and the orbital pair $(E,j)$. 

Finally, it is worth mentioning that it is of considerable convenience to use a rescaling for the
mechanical energy too. The varying density slopes are in fact responsible for strong variations in the central depth of the (dimensionless) potential.
This makes values of the energy very difficult to interpret in general, and would result in strong degeneracies in any parameter space exploration
covering a range of density profiles. For this reason, I adopt the mapping
\begin{equation}
{E\over{G \rho_0 r_0^2} }\rightarrow {r_{c}\over r_0} \ ,
\label{Etorc}
\end{equation}  
which univocally associates the mechanical energy $E$ with the radius $r_c$ of the circular orbit having that energy in the 
potential defined by the pair $(\gamma_i,\gamma_o)$. Together with the exponents~(\ref{slopes}), the circularity $j\in[0,1]$,
the time-associated dependences
of eqn.~(\ref{aphase}),~(\ref{atime}) and~(\ref{aradius}), the scaling~(\ref{Etorc}) grants that 
all free variables in the library are confined to conveniently bound and manageable intervals,
\begin{equation}
\log_{10}\left(r_c\over r_0\right)\in[-2,2]\ .
\label{radii}
\end{equation}
\begin{figure*}
\centering
\hspace{.03\textwidth}\includegraphics[width=.3\textwidth]{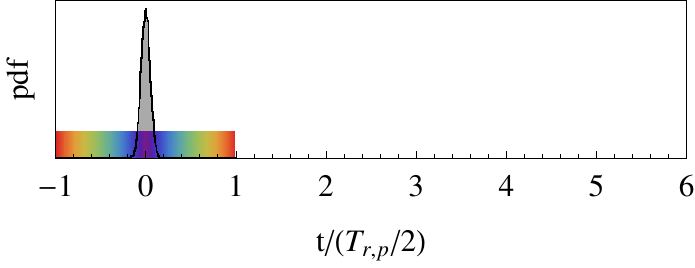}
\hspace{.03\textwidth}\includegraphics[width=.3\textwidth]{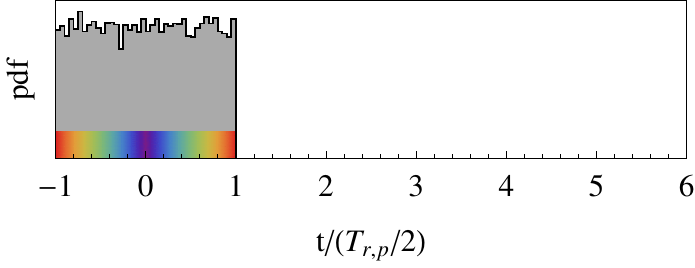}
\hspace{.03\textwidth}\includegraphics[width=.3\textwidth]{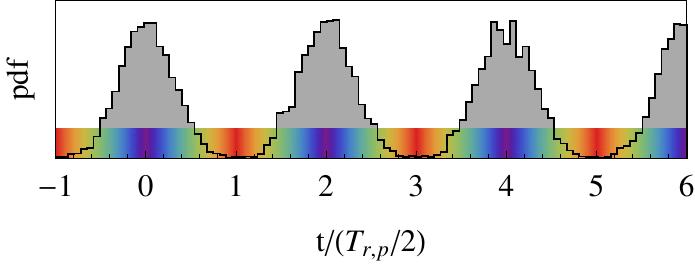}\\
\includegraphics[width=.3275\textwidth]{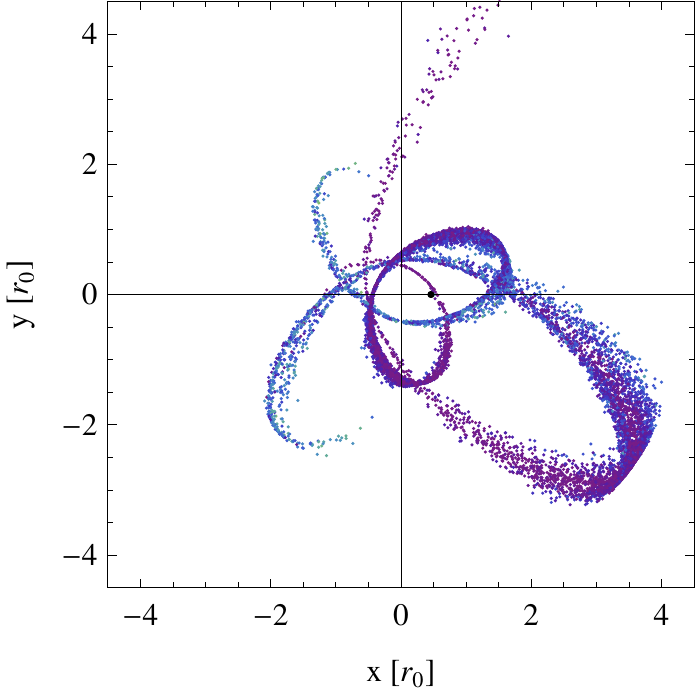}
\includegraphics[width=.3275\textwidth]{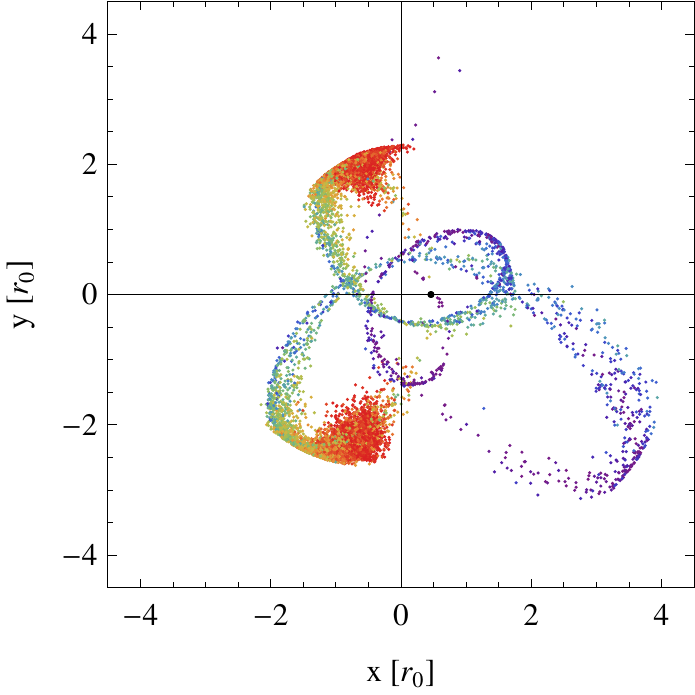}
\includegraphics[width=.3275\textwidth]{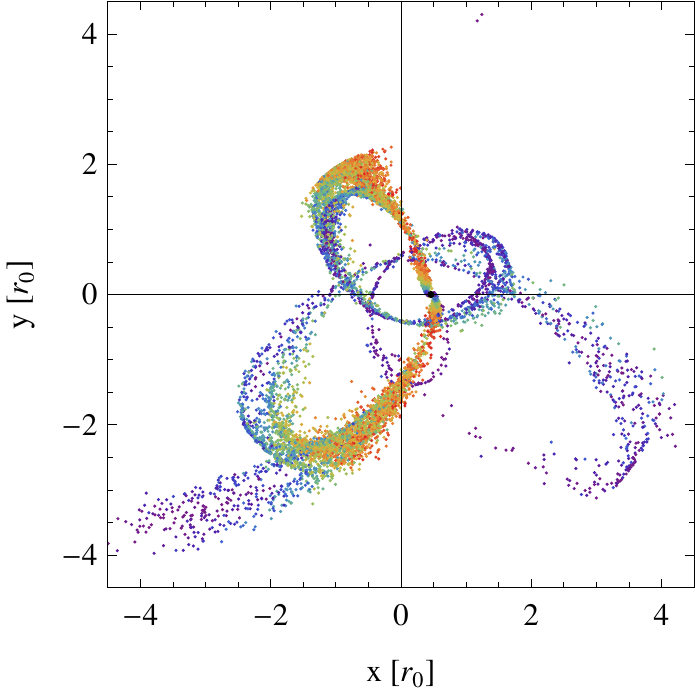}\\
\vspace{-.01\textwidth}
\includegraphics[width=.33\textwidth]{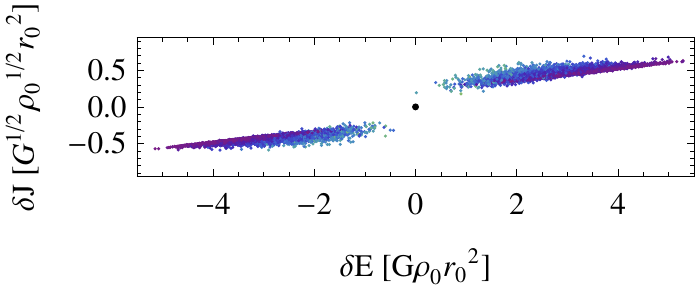}
\includegraphics[width=.33\textwidth]{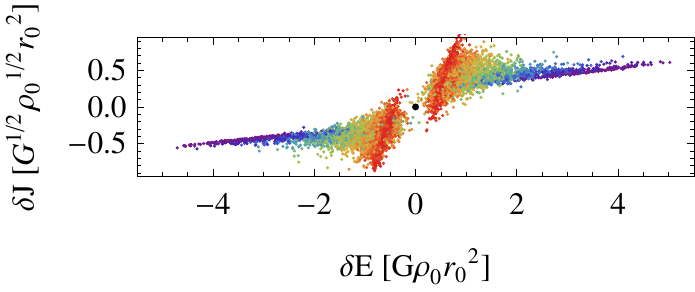}
\includegraphics[width=.33\textwidth]{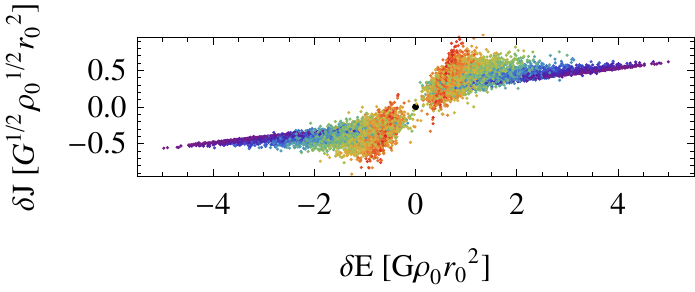}\\
\caption{The left, central and right panels compare the effects of different shedding histories, all other properties of the streams being the same.
Upper panels display the probability distribution functions of the shedding times, as a function of the normalised time $t/(T_{\rm r,p}/2)$, 
where $t=0$ is the first orbital pericenter. The stream is observed at $t=3 T_{\rm r,p}$, the fourth pericentric passage of the progenitor, which is shown as a black dot.
The lower panels show the distribution in energy and angular momentum of the leading and trailing members of the stream. The colour-coding is associated to 
the orbital phase of the progenitor at the time of escape, as indicated by the coloured bars in the uppermost panels.
{\it Left panels} illustrate the result of a single, almost impulsive shedding event, in which all material is dispersed at the first pericenter only,
with very little spread, $\sigma_{\bar t}=0.05$.
{\it Central panels} display the case in which shedding times have a uniform probability around the first pericentric passage, $\sigma_{\bar t}\gg1$, for a total of one orbital time since infall.
{\it Right panels} refer to the more realistic case in which the progenitor sheds material at each pericentric passage, with an intermediate spread of $\sigma_{\bar t}=0.3$ around each pericenter itself.}
\label{timepdff}
\end{figure*}

\subsection{Generation of members}

Given the stored orbital properties, the stream is constructed by generating its members.
For each single shed star, this procedure involves: 
(i) the generation of an escape time $t_s$;
(ii) the generation of a phase space displacement $(\bm{\delta r},\bm{\delta v})$;
(iii) the evolution in time given the previous two ingredients.
All members are released along the progenitor's orbit in the vicinity of the instantaneous tidal radius, 
in either the leading or trailing conditions~(\ref{rprime}). Eqn.~(\ref{rt}) fixes the modulation
of the tidal radius' magnitude with orbital time, while the size of the tidal radius at pericenter
$r_{{\rm t}, peri}$ is considered a free parameter of the model, approximately fixing the scaling $m/M$.
  
\subsubsection{Varying the shedding history}
Escape times are sampled along the progenitor's orbit according to a parametric probability distribution function (pdf).
This is modulated with pericentric distance, and hence with the normalised orbital phase and time $\bar t$.
A simple useful choice is a Gaussian pdf
\begin{equation}
{\rm p}(\bar t)={1\over{\sqrt{2\pi\sigma_{\bar t}^2}}}\exp\left[{-{1\over 2}\left({{\bar t-\langle\bar t\rangle}\over{\sigma_{\bar t}}}\right)^2}\right]\ .
\label{timepdf}
\end{equation}  
This allows me to set a delay of the shedding peak with respect to pericenter, through $\langle\bar t\rangle$, together with the 
`width' of the shedding episode around such peak, through $\sigma_{\bar t}$. Of course, more elaborated
functional forms can be used if they're found to best describe numerical simulations. 

Within the gaussian working hypothesis, the evaporative 
conditions defined in Sect.~2.2 are obtained by imposing that $\sigma_{\bar t}\gtrsim1$, which practically smooths out any orbital modulation and
determines an approximately constant shedding rate. The number of shedding episodes can also be fixed as desired, which
allows us to address a case in which the progenitor has been shedding tails at each pericentric passage since infall, or in which it 
has been virtually destroyed completely after a few orbits.

Figure~\ref{timepdff} shows an example of the qualitative differences in a stream's morphologies that are consequence
of changes in the shedding history. For each of the three columns, the upper panel displays the adopted pdf for the shedding times,
together with the colour-coding adopted in the
entire Figure. The first column displays the case of an almost impulsive shedding event ($\sigma_{\bar t}\ll1$) at the first pericenter 
$t =0$. The middle panels illustrate the opposite evaporative case of uniform release of particles for an entire orbital period, 
around the first pericenter only. The right column shows the more realistic case of repeated shedding events with an 
intermediate time spread ($\sigma_{\bar t}=0.3$). In all three instances all other physical parameters are kept fixed,
streams are populated with $10^4$ members and observed at $t/T_{r,p}=6$, the fourth pericentric passage of the progenitor (displayed as a black dot),
while the progenitor mass is kept constant with time. A few noteworthy differences are particularly evident.
\begin{itemize}
\item{Although the rosette configuration of the three streams is analogous, as orbit and potential are the same in all cases, 
the relative density of different regions of the stream varies substantially.
For instance, in the first two panels the tails generated at the first pericentric passage have completely detached from the progenitor.
Renewed shedding has taken care to fill those regions in the third column. }
\item{Similarly, the shifts in orbital phase between stars released at different passages introduce features in the third column 
that are completely absent from streams in the first two, as plumes at large radii.}
\item{From the third row of panels, it is clear that the energy and angular momentum differences $(\delta E, \delta J)$ in the typical bow-tie distribution
of the two tidal tails are strongly connected to shedding time. Larger differences in mechanical energy are achieved for stars released at pericenter, 
while stars released at apocenter have smaller $\delta E$ but a substantially larger spread in $\delta J$. As explored in Sect.~4 this is at the origin 
of both feathers and bifurcations.
}
\end{itemize}
\begin{figure}
\centering
\includegraphics[width=\columnwidth]{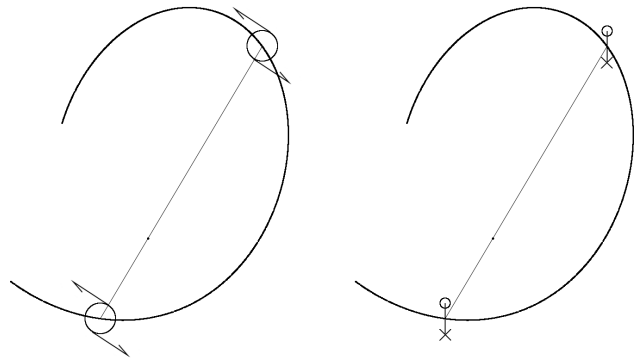}
\caption{Schematic representation of the systematic influence along the orbit of the progenitor's internal rotation on the escape velocities of leading and trailing particles. {\it Left panel}: rotation in the orbital plane contributes to a non-zero shift $\langle\delta v_{\varphi}\rangle$ of opposite sign for leading and trailing tails, with approximately constant sign and magnitude along the orbit. {\it Right panel}: a tumbling component of the rotation, as for example from an inclined disk, results in shifts $\langle\delta v_{z}\rangle$ that, for each tail, switch sign along the orbit as the orientation of the progenitor changes.}
\label{vscheme}
\end{figure}
\begin{figure}
\centering
\includegraphics[width=.4\textwidth]{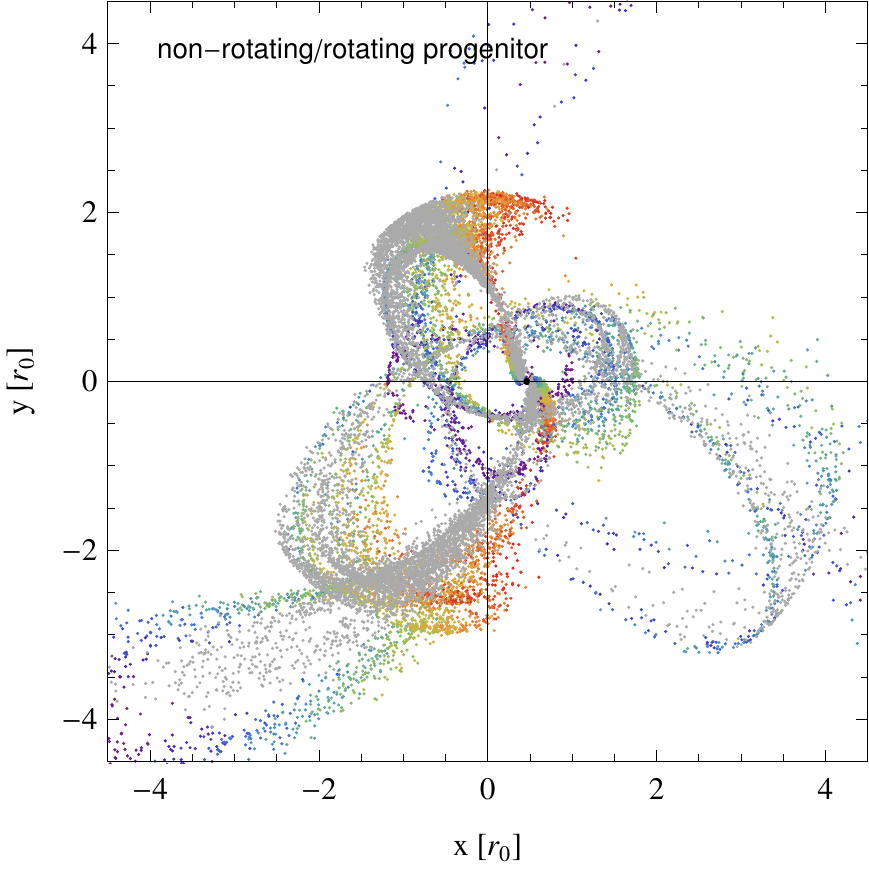}
\includegraphics[width=.41\textwidth]{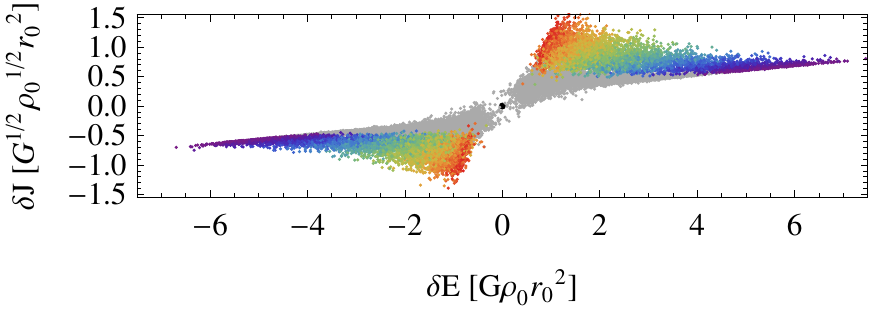}
\includegraphics[width=.38\textwidth]{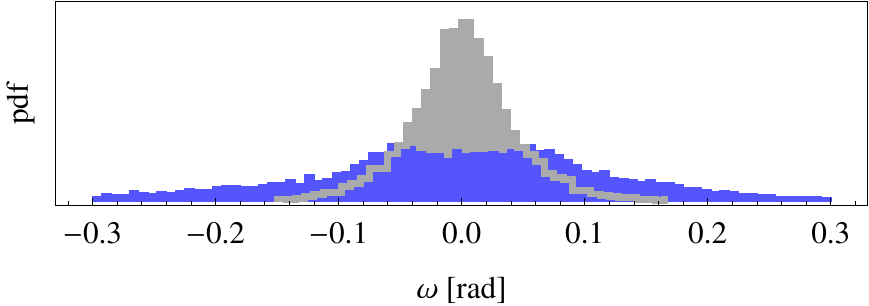}
\caption{The comparison between the stream produced by a progenitor having no internal rotation (in grey) and a progenitor that rotates in a prograde fashion with an angle
of $45^\circ$ with respect to the orbital plane. The non-rotating stream is the same as in the 
third column of Fig.~\ref{timepdff}, as is colour coding of the rotating stream; both are observed at the forth pericenter. {\it Upper panel}: 
the comparison between the distribution of members in the orbital plane. {\it Middle panel}: the comparative distribution in the $(\delta E, \delta J)$ space.
{\it Lower panel}: the distributions of the tilts of the orbital planes of the members (non rotating progenitor in grey, rotating one in blue).}
\label{rotcomp}
\end{figure}

\subsubsection{Varying the escape conditions: $\bm{\delta v}$}

At the zeroth order, the magnitude of the kinematic displacements $\bm{\delta v}$ in the velocities of shed stars 
is a measure of the progenitor's mass. However, as discussed in Sect.~2, such link between is not univocal, 
and involves the ratios $\rh/\rt$ and $\rh/\rp$. To maintain complete generality, this generative model 
considers the tidal radius at pericenter $r_{\rm t,peri}$ and the magnitude of the spread in the kick velocities at escape 
$\sigma_{\rm s}$ as two independent parameters of each shedding event.

At the first order, there are at least two reasons for which it is important to have flexibility not only in the 
magnitude of the kick velocities $\bm{\delta v}$, but also in the relative scalings of different components $(\delta v_r,\delta v_\varphi,\delta v_z)$.
\begin{itemize}
\item{First, as shown by \citet{GS14}, the distributions of $\delta v_r$ and $\delta v_\varphi$ are affected by the self gravity
of the progenitor when this is significant. In particular, self gravity is able to shift mean values away from zero, as only particles
moving in the direction of the Jacobi apertures will be able to escape \citep[see also][]{RF11}. This breaks the symmetry in the 
dispersion along different directions, introducing an approximate ordering 
\begin{equation}
\sigma(\delta v_r)\gtrsim\sigma(\delta v_\varphi)\gtrsim\sigma(\delta v_z)\ .
\label{selfg}
\end{equation}
}
\item{Second, the internal kinematics of the progenitor is able to introduce asymmetries in the distributions
of the kick velocities of leading and trailing arm. As a consequence of internal rotation in the progenitor, the distributions of the kick
velocities $\delta v_\varphi$ and $\delta v_z$ may instantaneously have non-zero means of opposite sign for particles that escape 
from the leading or trailing tidal radii.}
\end{itemize}

Given the above, I choose to parametrize the probability distribution functions of the different components of the kinematic displacements $\bm{\delta v}$
with three independent Gaussians, with explicit dependences on the associated tidal arm. In such a way it is possible to mimic 
the disruption of more massive satellites and of rotating progenitors, thereby expanding the applicability of this model. 

In the radial direction, particles escape more easily if their velocity points away from the progenitor, which implies shifts of opposite sign for the 
leading/trailing escape conditions:
\begin{equation}
{\rm p}(\delta v_{r})={1\over{\sqrt{2\pi\sigma_{\rm s,r}^2}}}\exp\left[{-{1\over 2}\left({{\delta v_{r}-\st\langle\delta v_{r}\rangle}\over{\sigma_{{\rm s},r}}}\right)^2}\right]\ ,
\label{velpdfr}
\end{equation}  
where $\st\in\{-1, +1\}$. 
In the azimuthal direction $\delta v_{\varphi}$, both self gravity and internal rotation might play their roles. However, for each tail, both of them contribute 
to an approximately constant shift of the mean of the pdf. Therefore, a single parameter can be adopted to describe 
them at the same time, using the same parametrisation as in eqn.~(\ref{velpdfr}):
\begin{equation}
{\rm p}(\delta v_{\varphi})={1\over{\sqrt{2\pi\sigma_{\rm s,\varphi}^2}}}\exp\left[{-{1\over 2}\left({{\delta v_{\varphi}-{\st}\langle\delta v_{\varphi}\rangle}\over{\sigma_{\rm s,\varphi}}}\right)^2}\right]\ .
\label{velpdfp}
\end{equation}  
Things are different for the kicks in the direction perpendicular to the progenitor's orbital plane $\delta v_{z}$. If the progenitor's internal angular momentum is not 
aligned with the orbital one, the mean $\langle \delta v_{z}\rangle$ changes in time at each Lagrange point. This modulation reflects the varying orientation of
the direction $\bm{\hat\rp}$ and the direction of the internal angular momentum $\bm{J_{\rm int}}$ of the progenitor. As made clear by the schematics of Fig.~\ref{vscheme},
it is easy to see that, while 
\begin{equation}
 \bm{\hat\varphi}\cdot(\bm{\hat{r}_{\rm t}}\wedge\bm{\hat{J}_{\rm int}})
\label{vphiconst}
\end{equation}  
is a constant along the orbit, 
\begin{equation}
 \bm{\hat z}\cdot(\bm{\hat{r}_{\rm t}}\wedge\bm{\hat{J}_{\rm int}})
\label{vzchange}
\end{equation}  
changes sign when the direction $\bm{\hat\varphi}$ is rotated by an angle $\pi$, {as also shown by \citet{PJ10}}. 
This dependence can be phenomenologically captured using the parametrisation
\begin{equation}
{\rm p}(\delta v_{z})={1\over{\sqrt{2\pi\sigma_{{\rm s},z}^2}}}\exp\left[{-{1\over 2}\left({{\delta v_{z}-{\st}\cos\left[\varphi_{\rm p}(t)-\varphi_0\right]\langle\delta v_{z}\rangle}\over{\sigma_{{\rm s},z}}}\right)^2}\right]\ .
\label{velpdfz}
\end{equation}  
This is identical to eqns.~(\ref{velpdfr}) and~(\ref{velpdfp}) apart for a cosine modulation. 
This gradually flips the mean kick velocity in the direction perpendicular to the orbital plane,
mimicking the periodic change in orientation of the progenitor. The phase $\varphi_0$ 
sets the initial orientation.

In conclusion, this framework mimics a rotating progenitor by varying the two dimensionless ratios  
\begin{equation}
\left({{\langle\delta v_{\varphi}\rangle}\over{\sigma_{\rm s,\varphi}}} , {{\langle\delta v_{z}\rangle}\over{\sigma_{{\rm s},z}}}\right)\ .
\label{rotshifts}
\end{equation}
As shown by \citet{GS14}, self gravity can be responsible of $\langle\delta v_{\varphi}\rangle>0$,  up to 
${{\langle\delta v_{\varphi}\rangle}/{\sigma_{\rm s,\varphi}}}\lesssim 1$, 
higher values can be used to mimic (prograde) rotation. 
Fig.~\ref{rotcomp} shows the comparison between the stream displayed in the third column of Fig.~\ref{timepdff} 
with a stream that is generated by keeping all parameters fixed, apart from the dimensionless shifts of eqn.~(\ref{rotshifts}).
The stream of Fig.~\ref{timepdff}, displayed in gray in Fig.~(\ref{rotcomp}), 
has $(\langle\delta v_{\varphi}\rangle/\sigma_{\rm s,\varphi}, \langle\delta v_{z}\rangle/\sigma_{\rm s,z})=(0.5,0)$,
as self-gravity could cause, hence corresponding to a non rotating progenitor. 
The stream it is compared to has $(\langle\delta v_{\varphi}\rangle/\sigma_{\rm s,\varphi}, \langle\delta v_{z}\rangle/\sigma_{\rm s,z})=(4,4)$,
which corresponds to a purposely strong prograde rotation, nominally inclined by $\approx45^\circ$ with respect to the orbital plane.
To simplify the comparison, the colour-coding used in Fig.~\ref{rotcomp} is the same of Fig~\ref{timepdff}. 

The stream generated by the rotating progenitor has a faster 
differential streaming, caused by its larger extension in $\delta E$ in the bow-tie plot.
The rotating progenitor also results in a much richer stream, with features that are globally more marked and 
better defined. 
The inclined rotation causes the stream to loose its purely planar nature, as the nonzero shifts in $\delta v_z$ cause significant tilts in the orbital
planes of the stream members. The lower panel of Fig.~\ref{rotcomp}
compares the distributions of such tilts $\omega$. Note in particular that when internal rotation is tilted 
with respect to the orbital plane, each single arm loses its planar nature despite 
the spherical symmetry of the potential, as systematic variations of the orbital tilts $\omega$ with time are introduced. 
This may be useful to explain the non planar nature of Sagittarius' bifurcation.

Finally, it is fair to say that internal rotation can be responsible for a much wider range of complex phenomenologies,
arising for example through the excitation of resonances \citep{DE10}, and that cannot be easily included in this simple model. 
However, resonances are only excited when the internal frequencies of the progenitor and the orbital frequencies of the host are comparable. 
This happens for mass ratios $m/M$ at the upper edge of the interval we are interested in, where self gravity and dynamical friction are bound 
to introduce even more worrying threats to the present model. 

\subsubsection{Varying the escape conditions: $\bm{\delta r}$}

{ As discussed in Sect.~2, stars escape from apertures in the surface of the effective
potential. In Sects.~2.2 and~2.3 I have shown that the comparison between their size $\varpi$,
and the nominal tidal radius $\rt$ or pericentric distance $\rp$, drives the global morphology
of the stream. As a consequence, it is important to be able to vary such characteristic 
size in the generative model. 

Stars leave the progenitor from the vicinity of the Lagrange points but the 
size of these apertures depend on the energy of the escaping particles, as calculated explicitly in
eqns.~(\ref{semiax1}) and~(\ref{semiax2}). I will use the mentioned scalings to set the 
escape conditions of each stream member: given a velocity kick of magnitude $\delta v$,
eqns.~(\ref{semiax1}) and~(\ref{semiax2}) set the characteristic scales of the corresponding
escape region. For simplicity, I adopt a simple Gaussian parametrisation, so that
\begin{align}
{\rm p}(\varpi_r)&=\delta(\varpi_r)\ ,\\
{\rm p}(\varpi_\varphi)&={1\over{\sqrt{2\pi\bar\varpi_{\varphi}^2}}}\exp\left[{-{1\over 2}\left({{\varpi_\varphi}\over{\bar\varpi_{\varphi}}}\right)^2}\right]\ ,\\
{\rm p}(\varpi_z)&={1\over{\sqrt{2\pi\bar\varpi_{z}^2}}}\exp\left[{-{1\over 2}\left({{\varpi_z}\over{\bar\varpi_{z}}}\right)^2}\right]\ ,
\end{align}  
where $\delta$ is here the usual Kronecker delta function, and the scales $\bar\varpi$ are
intended as functions of $\delta v$,  .
This allows me to set the scale $\sigma(\varpi)$ of the escape regions,
and therefore to address all regimes studied in Sect.~2.}

\section[]{Coherent streams: Feathers and bifurcations }

Given the analytical and numerical frameworks described above, in this Section I deepen on the dynamics 
and properties of streams that belong to the following two categories 
\begin{itemize}
\item{feathers: slow and coherent streamers, as defined respectively by eqn.~(\ref{slowc}) and eqn.~(\ref{evap}) or~(\ref{evap2});}
\item{bifurcations: fast and coherent streamers, as defined by eqn.~(\ref{fastc}) and eqn.~(\ref{evap}) or~(\ref{evap2});.}
\end{itemize}
As discussed in Sect.~2.2, internal coherence is the condition to the formation of substructure in the stream.
The properties of overdensities and gaps in the tails of disrupting star clusters has
been recently studied by several authors \citep[][]{CD05, Ju09, KA10, KA12, DE14}. 
Part of the interest sparks form the consideration that disturbances in these patterns 
may represent a promising venue to quantify the population of dark matter subhaloes 
of the MW. Preliminary studies have been performed using mainly streams shed by 
GCs \citep{Yo11, Ca12, Ca13, NW14}, but is some cases more massive progenitors 
have also been considered \citep{SG08}.

\begin{figure*}
\includegraphics[width=.44\textwidth]{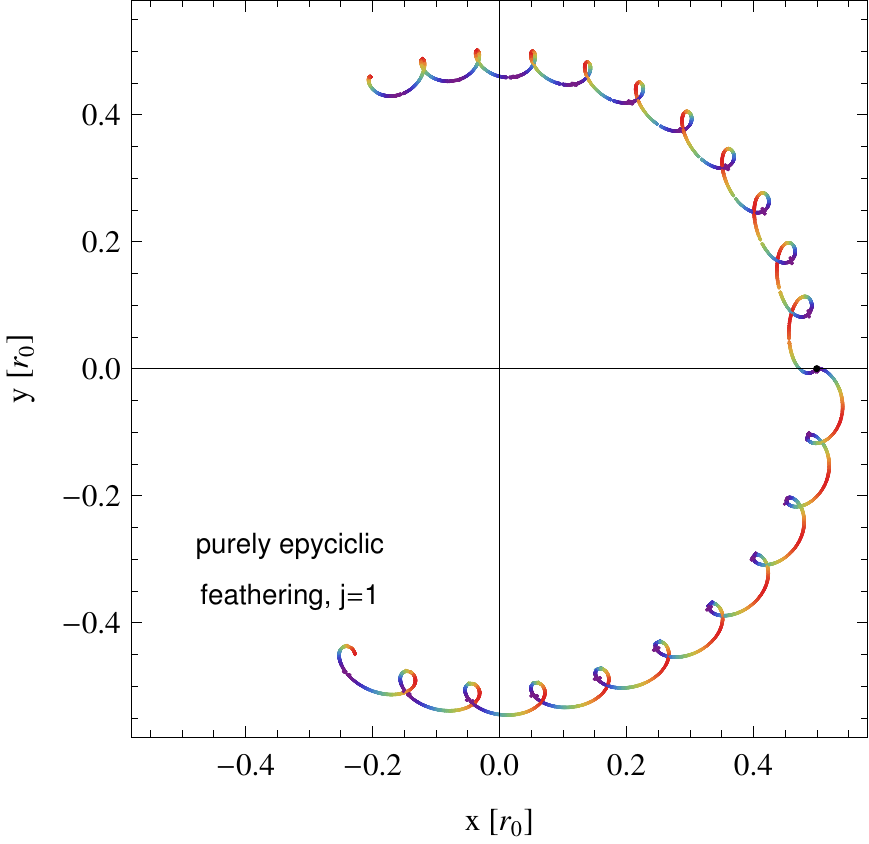}
\includegraphics[width=.44\textwidth]{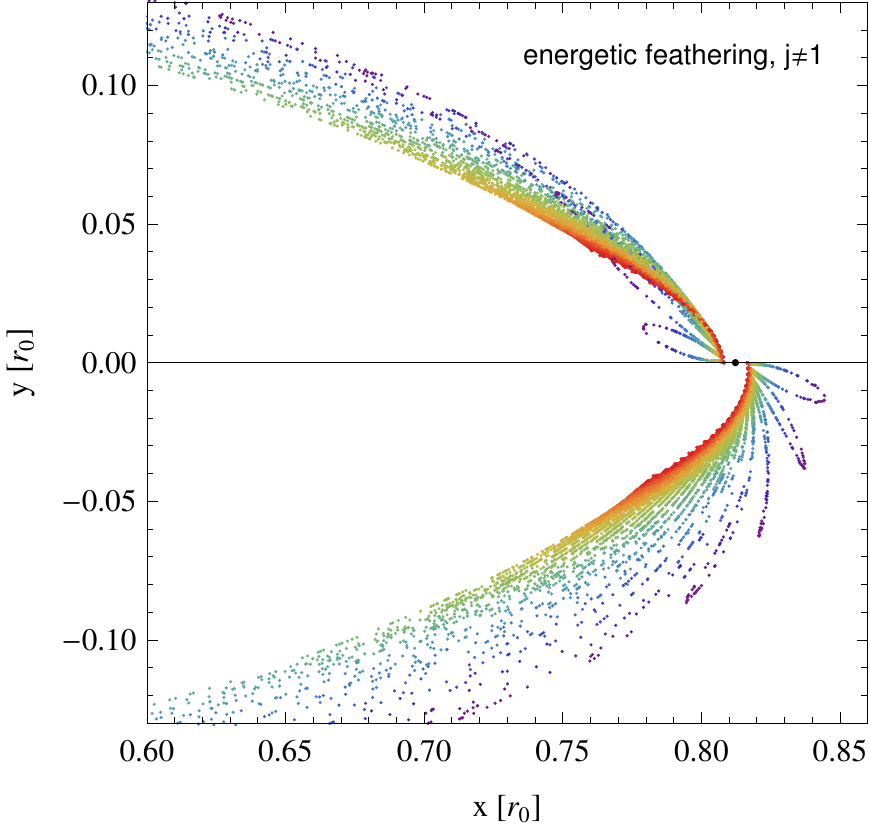}
\includegraphics[width=.34\textwidth,angle=90]{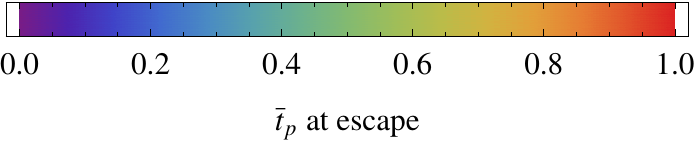}\\
\includegraphics[width=.44\textwidth]{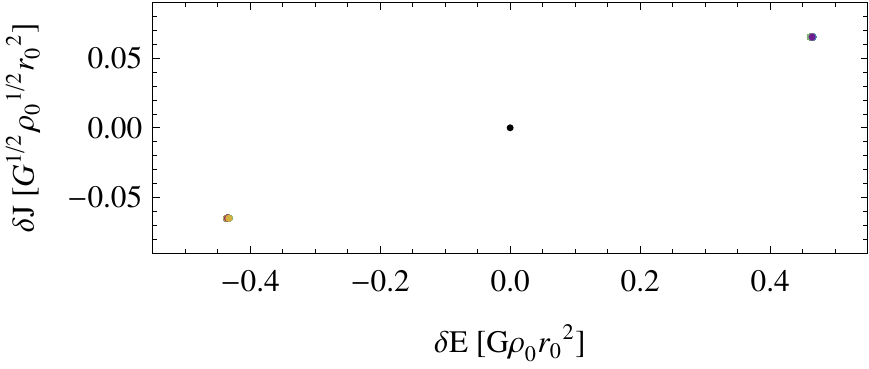}
\includegraphics[width=.44\textwidth]{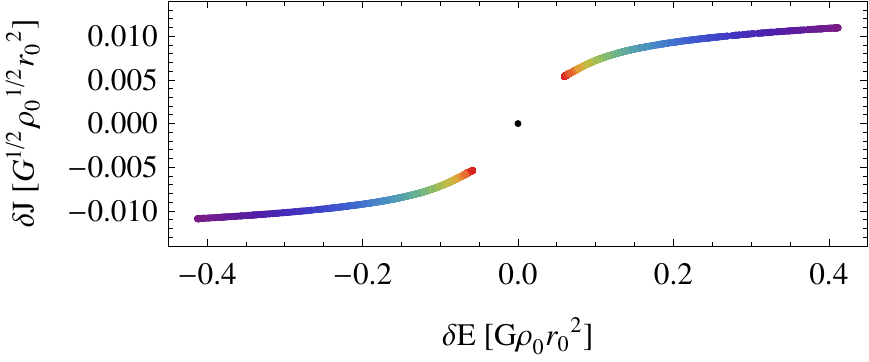}
\hspace{1.3cm}
\caption{Two streaklines illustrating the generation of purely epicyclic feathers (left panels) in the case of a circular orbit $\jp=1$, and of energetic feathers (right panels) in the general case $\jp\neq1$. The colour-coding displays the orbital phase of the progenitor at the time of escape, as shown by the colour scale on the right. $\bar t_{\rm p}=0$ indicates the orbital pericenter, $\bar t_{\rm p}=1$ the apocenter. In the general case $\jp\neq1$, the progenitor's phase at escape is in one-to-one correspondence with the phase space displacements $(\delta E,\delta J)$ (lower panels). This implies a one to one correspondence with the ordering of particles along the stream. Each armlet, composed of particles shed between two successive apocenters, is folded  along most of its length, pulled at its centre by the faster differential streaming of stars released at pericenter, with larger $|\delta E|$. This mechanism is secular, and therefore dominant on the purely epicyclic feathering of the left panels, in which particle positions are determined by the time passed since escape. Overdensities correspond to the local minima in the differential streaming speed of  members. These have different natures in the two cases: are local minima (evident as `ripples' in the streakline's geometry) at fixed locations with respect to the progenitor and crossed by different stars at different times if $\jp\approx1$; are associated with the lower streaming speeds of particles lost away from pericenter, they systematically stream away from the progenitor and are constantly composed of the same stars. }
\label{feathers}
\end{figure*}

\subsection{Slow streams: feathers}

The origin of substructures in stellar streams has been identified in the epicyclic motion of the escaped stars. Indeed, 
when the progenitor's orbit is perfectly circular, stars are released with $j_{\rm s}\approx1$,
a regime that is very well described by the epicyclic approximation. For example, \citet{KA12}
used this formalism to quantitatively reproduce the streaklines of star clusters orbiting with $j_{\rm p}=1$.
On the other hand, the general case $j_{\rm p}\neq 1$ is not tackled so easily in an analytical way. Using numerical methods,
\citet{KA12} were able to show the richness of morphologies that streaklines can achieve, and to tune them to 
reproduce the tracks of streams produced in N-body simulations. Here, I note that the different and richer phenomenology 
of the general case $\jp\neq1$ is caused by an different mechanism, due to a modulation with time of the energies $\delta E$ 
of the shed particles. 

A streakline illustrating the ideal case $\jp=1$ is illustrated in the left panels of Fig.~\ref{feathers}. 
Colour-coding displays the orbital phase of the progenitor at the time of escape of each particle, 
with $\bar t_{\rm p}=0$ being a pericenter. As it can be seen in the lower left panel, 
the escape conditions around the circular orbit determine constant shifts $(\delta E, \delta J)$, with no time dependence, as
galactocentric distance is constant. This implies that the angular distance of each particle from the progenitor grows at the same secular rate: 
relative positions along the stream are determined by escape time only. This also means that secular differential streaming is inhibited within the 
particles themselves, as all of them have been released with very similar energies, generating none of the secular components isolated in eqn.~(\ref{phid2}). 
However, because of the epicyclic oscillations, the angular distance between two particles belonging to the same tail
\begin{equation}
\varphi_{1,2}(t)= \varphi_{\rm s,1}(t)-\varphi_{\rm s,2}(t) 
\label{sangdiff}
\end{equation}
oscillates in time with the epicyclic frequency $\kappa$.
As shown in Fig.~\ref{feathers}, this implies `ripples' in the geometry of the streakline: in those locations differential angular speed
is temporarily lower. As local density anti-correlates with the speed of the relative differential streaming, this results in overdense regions. 
Each of these overdensities lies at fixed locations with respect to the progenitor, those locations reached in $t\approx2 n \pi/\kappa$. As instead 
stream members do systematically move away from the progenitor, each of these overdensities is composed of different stars at different times.

On the other hand, the right panels of Fig.~\ref{feathers} illustrate the general case $\jp\neq1$.
Because of the varying galactocentric distance, a one to one correspondence between the displacements $(\delta E, \delta J)$ and the
orbital phase of the progenitor $\bar t_{\rm p}$ is apparent: the term $\delta E_{\rm t}$ is a direct function of time,
modulated by the radial period $T_{\rm r,p}$. This correspondence is clearly mirrored in the angular distance of
particles from the progenitor: particles released near pericenter experience a faster differential streaming, through eqn.~(\ref{phid2}). 
Particles released near apocenter are instead forced to lag along each tidal tail. As this is true at the same time for particles released at the apocenters 
$ t_{peri}-T_{r,p}/2$ and $ t_{peri}+T_{r,p}/2$, the corresponding section of the streakline is folded along most of its length. 
This determines the appearance of feathers. 

Epicyclic overdensities are in this case suppressed: due to the modulation with galacticentric distance, 
the angular difference $\varphi_{1,2}$ has a secular component (unless particles 1 and 2 
have been released when the progenitor was at the same orbital phase), and particles systematically stream away from each other. 
Overdensities along an eccentric stream have therefore a different nature. 
If the escape rate is not exceedingly low at apocenter, particles released away from pericenter are more tightly packed 
as a result of their lower (absolute and relative) differential speed. As shown by the right panels of Fig.~\ref{feathers} in the form of (red) knots, they 
form overdense regions close to the progenitor. Note that these overdensities are intrinsically 
different from the ones generated by epicyclic oscillations: they are formed by the same stars at all times and they drift away from the progenitor 
(although with the low streaming speed of material shed at apocenter).


Not all generative models are equally capable of capturing the formation of feathers, as not all of them can account for the 
time-modulation that is at its origin. Of course, the `streakline-method' as presented by \citet{KA12}, the variation of \citet{GS14} 
and the one introduced in this paper capture this time dependence, as particles are released from a time dependent tidal radius. 
However, this is more difficult to achieve for methods that rely on the simple structure of streams 
in action-angle space. In this context, feathers are determined by a modulation of the frequency shifts (with a slight abuse of notation) 
$\bm{\delta\Omega}$ with time, or in other words with the progenitor's angle $\bm\theta_{\rm p}$ at the time of escape. The existence of such
modulation has been shown explicitly in very recent numerical work by \citet{FM14} and \citet{CR14}. 
If this time dependence of the actions on time is averaged out, like for example in \citet{BJ14}, feathering and
substructures are not observed. 

\begin{figure}
\centering
\includegraphics[width=.4\textwidth]{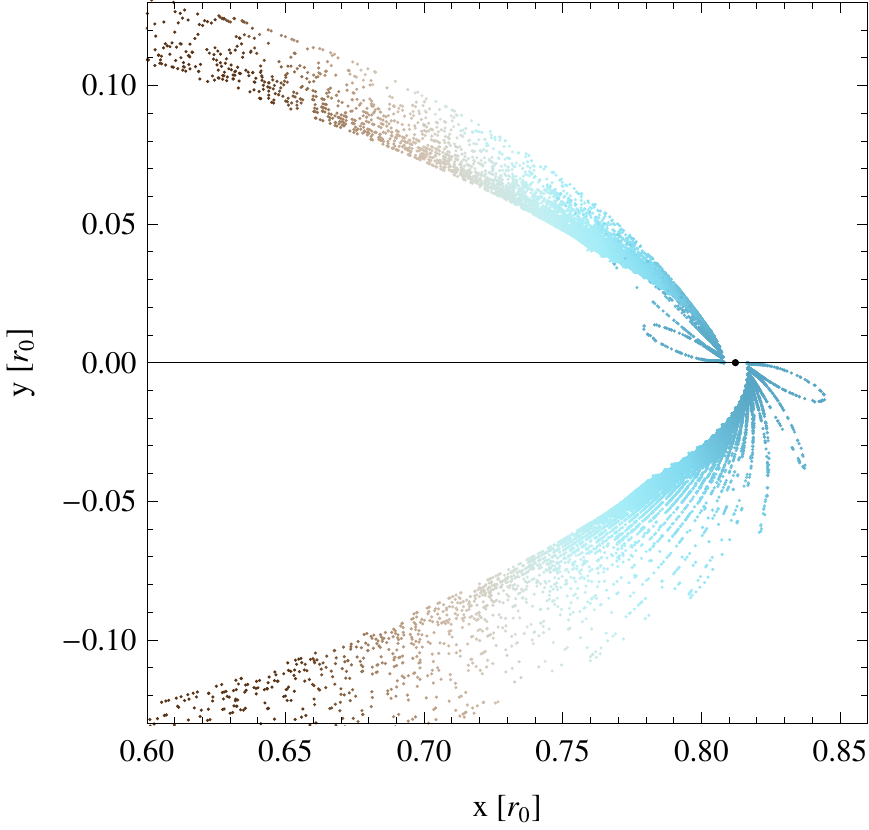}
\includegraphics[width=.34\textwidth]{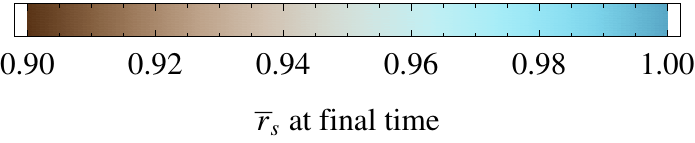}
\caption{The radial phase of particles in a feathered stream.  }
\label{radphase}
\end{figure}

Having determined the origin of the feathers allows me to address their size. For example, it is easy to quantify the radial size of an armlet 
at apocenter, where, as shown in Fig.~\ref{radphase}, we can assume that particles within the last shed armlet of a slow stream have a similar 
radial phase (note that the scale only ranges in the interval $\bar r\in[0.9,1]$). Therefore we can use the following approximation
\begin{equation}
r_{\rm s, apo}-r_{\rm p, apo}\approx \left[\bm{\nabla}_{(E,j)}r_{apo}\right]_{(\Ep,\jp)}\cdot \binom{\delta E}{\delta j}\ ,
\label{armletsize}
\end{equation}
where, as in eqn.~(\ref{phid}), the gradient indicates derivatives with respect to energy and circularity. 
We can capture the dependences of the dominant contribution to eqn.~(\ref{armletsize}) by considering a scale-free density profile. 
In such simple case, $r_{apo}(E,j)=r_{apo}(E,j=0)f(j)$; also, changes due to energy variations are dominant, so that
\begin{equation}
{{r_{\rm s, apo}-r_{\rm p, apo}}\over {f(j)}}\approx {{\partial r}\over{\partial \Phi}}\biggr|_{apo}\delta E\big|_{peri}
\approx {{\partial r}\over{\partial \Phi}}\biggr|_{apo}\ \left({{\partial \Phi}\over{\partial r}}\rt\right)\biggr|_{peri}\ .
\label{armletsize2}
\end{equation}
Eqn.~(\ref{armletsize2}) shows that the radial size of the armlets is strongly connected to the properties of the gravitational potential. 
As for eqn.~(\ref{dkappaasym}), both mass ratio $m/M$ and the steepness of the host's density profile play a relevant role. 
Note in particular that, if the properties of the host are known with some precision within the orbital pericenter, eqn.~(\ref{armletsize2}) shows 
that energetic feathering can be used to measure the steepness of the host's potential further out. For example, larger values of $\gamma_o$
are bound to result in measurably longer feathers. This link provides a surgical tool to probe the local properties of the host's density profile 
and for this reason, modelling techniques should aim at reproducing the entire phase space distribution of streams, including their detailed 
spatial density, rather than their averaged 'tracks'.

\begin{figure}
\centering
\includegraphics[width=.45\textwidth]{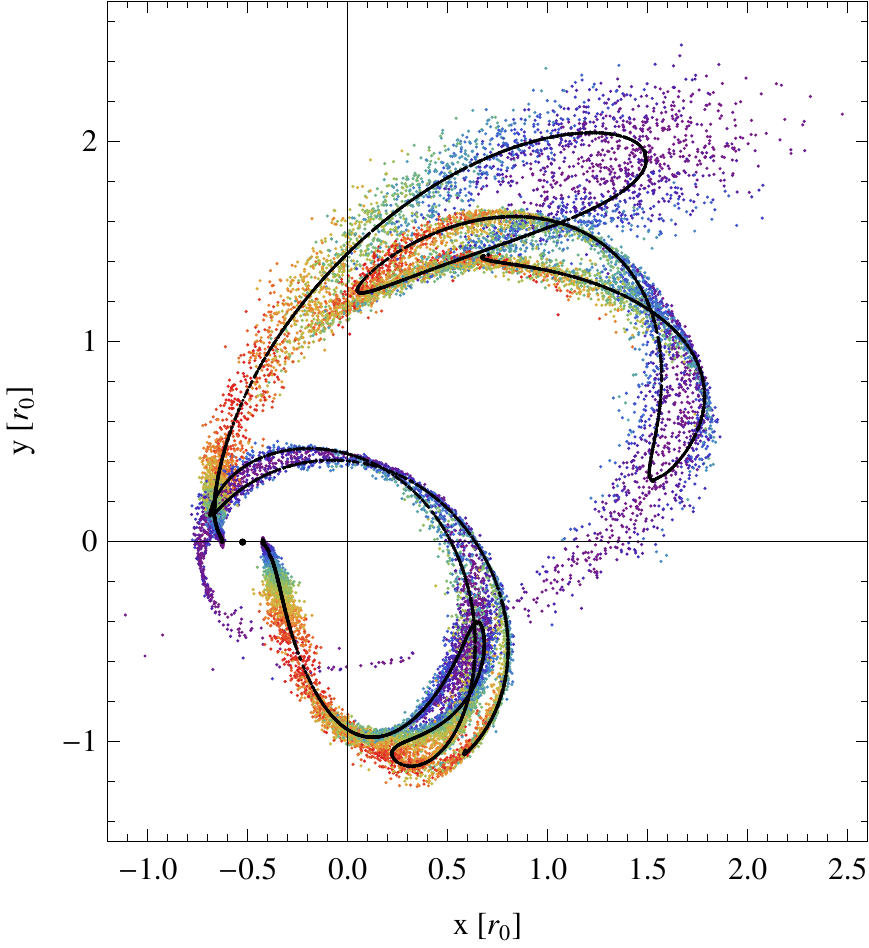}
\includegraphics[width=.34\textwidth]{timeleg-eps-converted-to.pdf}
\caption{A fast stream with coherent arms, displaying bifurcations. Particles released at pericenter $\bar t_{\rm p}=0$ experience a much faster
differential streaming than particles released at apocenter $\bar t_{\rm p}=1$ and each armlet is folded along most of its length, the associated streakline is shown in black. 
The stream here is observed at the third pericentric passage since infall, and the velocity dispersion of particles at escape is low enough to preserve 
internal coherence, allowing its bifurcated appearance to survive. }
\label{sagittarius}
\end{figure}

\subsection{Fast streams: bifurcations}

The previous Section has illustrated the formation of feathers in slow streams, in the sense of eqn.~(\ref{slowc}). Feathers appear as 
a collection of approximately parallel folded armlets, shed around successive pericentric passages. However, 
the energetic folding of armlets is a mechanism that does not depend on how fast the average differential streaming is and
it also implies the internal folding of the streaklines of long arms in fast streams. In other words, whatever the 
average magnitude of $\langle\delta E_{\rm k}\rangle$ across the leading or trailing tail, as 
 $\delta E_{\rm k}$ has a periodic modulation with time, the inner structure of each each armlet is bound to be intrinsically folded. 
Therefore, if the release conditions are cold enough, this mechanism can potentially cause an intrinsic bifurcation in the density of a tidal arm, 
without requiring multiple shedding events at different pericentric passages.

 For example, Fig.~\ref{sagittarius} shows the tails of a progenitor with mass $m/M\approx 7\times10^{-3}$ at pericenter,
 just after the third pericenter passage. The potential is not too different from logarithmic, with $(\gamma_i, \gamma_o)=(2,2.5)$. The mass of the
 satellite (and the potential steepness) are high enough that the stream is fast, and the tails encircle the host's centre after
 just 2.5 radial periods. The black lines show the streakline of each arm, displaying the typical folding. The two armlets contributed 
 from each of the two pericentric passages are pulled forward by particles shed at pericenter (in purple, colour coding is the same as in Fig.~\ref{timepdff},~\ref{rotcomp} 
 and~\ref{feathers}) resulting in the folding. Both leading and trailing armlets shed during the first radial period since infall show a 
 very marked bifurcation. Random motions at escape have been fixed so that
 \begin{equation} 
 \sigma_{\rm s}\approx{1\over 3} \left[ \vp\left({m/ M}\right)^{{1\over3}}\right]_{peri}
 \end{equation}
 which, as prescribed by eqn.~(\ref{disrupt2}), allows the bimodality to survive despite some non negligible internal mixing.

\begin{figure}
\centering
\includegraphics[width=.4\textwidth]{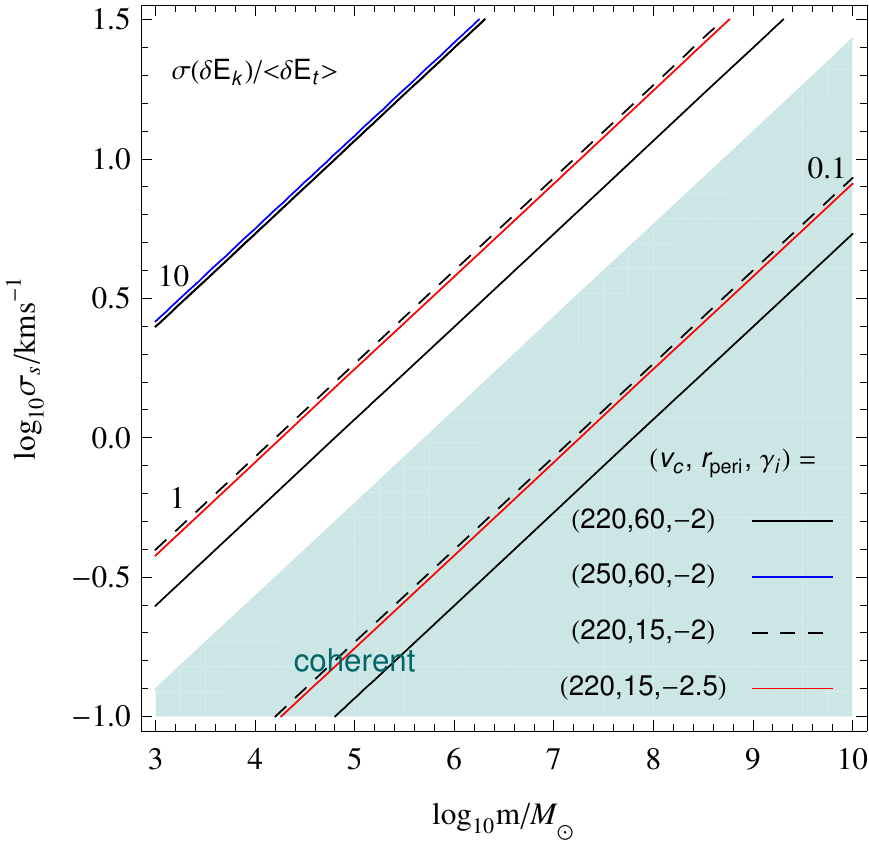}
\caption{An order-of-magnitude estimate for the regime of coherent streams around the Milky Way, where substructures, multiple tails and bifurcations are
possible. The structural parameters of the Galaxy are varied as in the legend, where $v_c$ is the circular velocity 
at the Sun, expressed in kms$^{-1}$, and a logarithmic potential ($\gamma_i=2$) is assumed as a reference. As shown by the different 
colours, reasonable variations of the structural parameters of the Galaxy are not influent. The effect of varying pericentric 
distances (expressed in kpc) are also explored. The green shading indicates the region where ${ {\sigma(\delta E_{\rm k})} / {\delta E_{\rm t}} } \lesssim 0.5$, which can be approximately associated with coherent streams.}
\label{cohef}
\end{figure}

\section{Coherent tails across the mass scales}

Feathers and bifurcations are apparent when the stream is internally 
coherent, in the sense of eqn.~(\ref{evap}). In the opposite case of hot escape conditions, random motions internal to the stream
wash out the bifurcation, as shown for example by the middle panel of Fig.~\ref{regimex}. However, in real cases, it is not straightforward 
to classify the level of coherence of a stream. 
As derived in Sect.~2.2, this is determined by the relation between the 
distribution of energies in the progenitor and the critical Jacobi constant $\bar{E}_{\rm J}$. 
In order of magnitude, the ratio $\rh/\rt$ is a good indicator: if
$\rh\ll\rt$, then eqn.~(\ref{evap2}) is satisfied and the stream is coherent;
in the opposite case of eqn.~(\ref{hot0}) particles stream freely outside the tidal radius and tails are warm in the sense of eqns.~(\ref{hot2}).

The analysis of Sect.~2.2 can be used to roughly explore the interplay between the physical 
ingredients that determine the appearance/disappearance of internal substructure. Assuming the Milky Way 
as the host, Fig.~\ref{cohef} shows the contours of the dimensionless ratio considered in eqn.~(\ref{disrupt3}):
 \begin{equation}
\epsilon\equiv{ {\sigma(\delta E)} \over {\delta E} } \approx {\sigma_s\over\vp}\left[m\over M(r_{peri})\right]^{- {1\over 3}}\ .
 \label{cohe}
 \end{equation}
The progenitor mass $m$ and the spread in the distribution of kick velocities $\sigma_{\rm s}$ are free, to allow for comparisons with real cases.
The host potential is assumed to be logarithmic ($\gamma_i=\gamma_o=2$) with a circular velocity $v_c$ between 220 and 250 kms$^{-1}$ --
variations in this interval are found to have no effect -- for simplicity, we assume $\vp\approx v_c$. Also, the influence of a more strongly declining 
density profile is explored, although its effect is also found unimportant here. The green shaded area in Fig.~\ref{cohef} identifies the region where 
${ {\sigma(\delta E_{\rm k})} / {\delta E_{\rm t}} } \lesssim 0.5$, which can be approximately associated with coherent streams. Larger spreads in 
the escape velocities are likely capable of washing out internal substructure.

\subsection{Palomar~5}

The internal coherence of the long tails of the GC Pal~5 is testified by their substructures, observed using SDSS data
\citep{Gr06d, Ca12}. Although the spread in the escape velocities is unknown, the internal kinematics of remnant's core 
has been probed by \citet{Od02, Od09}, which found a central velocity dispersion as low as $\sigma_{\rm p}\lesssim1$ kms$^{-1}$. 
Using photometric star counts, the same authors have also estimated that the total mass of the remnant is $m\approx5 \times 10^3\ M_{\odot}$.  
Assuming this estimate is correct, Fig.~\ref{cohef} confirms that the velocity dispersion of the remnant is in the correct range. 
In fact, in order to allow for $\epsilon\lesssim0.5$, the kinematic spread at escape should be even smaller, of just a fraction of a kms$^{-1}$.

\subsection{Willman~1}

\citet{WB06} noticed features that appear as `multiple tidal tails' around the ultrafaint Willman~I. 
Despite the low luminosity of the dwarf and its unfortunate systemic heliocentric velocity, mixing
foreground and members, \citet{WB11} were able to obtain an estimate of the internal velocity dispersion of the remnant, $\sigma_{\rm p}=4\pm0.8$ kms$^{-1}$.
Using this measurement, the mass of Willman~I is estimated by the same authors at $m\approx4\times 10^5\ M_{\odot}$, although, 
as the dynamical state of the remnant is not known, such face value is significantly uncertain.

If we assume that the features detected in the photometric number counts by \citet{WB06} are indeed the roots 
different armlets in a feather-pattern, we can use Fig.~\ref{cohef} to obtain a lower limit for the mass
of Willman~1. Assuming for example that $\sigma_s\approx1$ kms$^{-1}$, through $\epsilon\lesssim0.5$ Fig.~\ref{cohef} implies that the mass 
of Willman~1 is $\gtrsim10^6 \ M_{\odot}$, which confirms its galactic origin.

\begin{figure}
\centering
\includegraphics[width=\columnwidth]{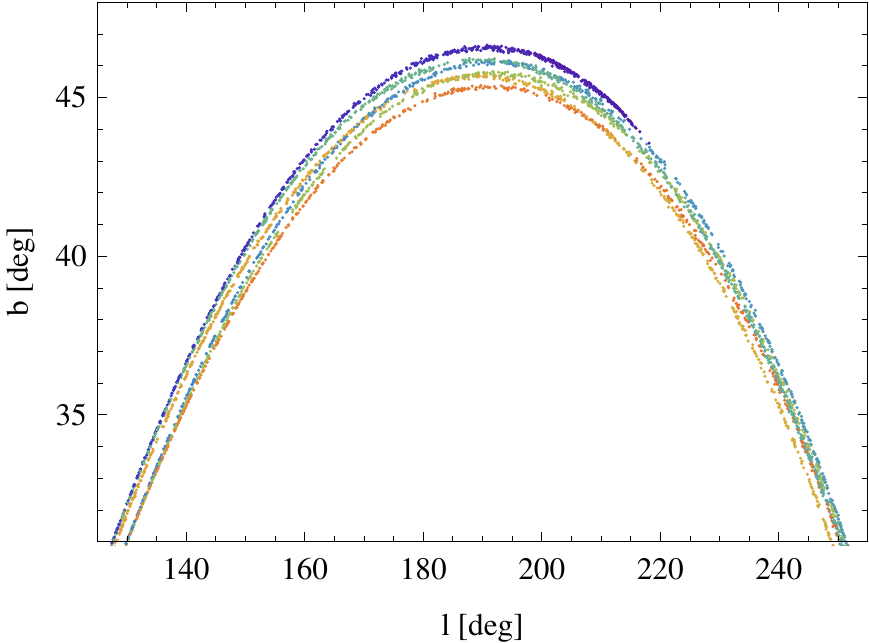}
\includegraphics[width=.8\columnwidth]{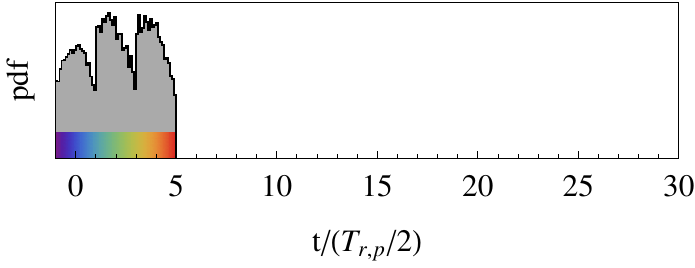}
\caption{A qualitative model displaying parallel overdensities similar to the internal structure of the Anticenter stream.
Material in the stream is colour-coded according to the time of its escape, during the first three orbital periods. The stream,
however, is observed after substantial differential streaming, when its slow tails have had enough time to encircle most
of the Galaxy.}
\label{anticenter}
\end{figure}
\subsection{Anticenter stream}

Among the most puzzling streams for its rich internal structure is the so called Anticenter stream \citep{Gr06c}.
This is relatively nearby, at only $\approx9$ kpc from the Sun, and is characterised by a number of parallel overdensities,
which, with varying intensities, sweep the entire length accessible in the SDSS data. The 
different density peaks have very similar distances, which brought \citet{Gr06c} to suggest that they may
have been caused by the disruption of the original GC population of the progenitor. Better constraints on the orbit 
have been obtained by \citet{Ca10}, though the origin of the internal substructure has so far remained elusive.

Without aiming to provide a complete model of the Anticenter stream, I note here that its multiple parallel overdensities 
do not necessarily require a substructured progenitor, and can in fact be interpreted as due to the natural internal
structure of a feather. After substantial differential streaming, the multiple armlets of a coherent stream naturally
appear as long, thin overdensities that run parallel to each other. These are shifted by a small angle as a result of having been 
shed ad successive orbital times. In this scenario, the progenitor of the Anticenter stream should have been destroyed after a few 
pericentric passages, giving rise to the observed parallel features.

Figure~\ref{anticenter} shows the qualitative morphology of a feather when observed after substantial differential
streaming. With the mass characteristic of an ultrafaint ($m/M\approx10^{-4}$ at pericenter), the tails of this progenitor 
are nominally slow in a logarithmic potential, but when observed after more than 10 pericentric passages have almost 
encircled the Galaxy. Material composing the 
stream has been shed during the first three orbital periods, resulting in a stream with a few thin parallel substructures,
which mimic the appearance of the Anticenter stream.

\subsection{Sagittarius}

The details of Sagittarius' history are still enigmatic despite considerable effort in both theory and observations. 
In particular, a mechanism capable of explaining the clear bimodality in the density of both leading and trailing 
arms \citep{BV06c, KS12} is yet to be identified. 

\citet{Fe06} propose that the apparent bifurcation is in fact an artificial effect, produced by the presence of two different 
wraps shed at successive pericentric passages. Although at similar distances, the two armlets are slightly shifted with respect to each other, 
leading to the detection of what appears as a bifurcation. This interpretation 
is able to account for the different chemical properties of the two parallel density peaks \citep[see e.g.,][and references therein]{Ch07, Sh12, Cas12}, 
as these are made of material stripped from ever more central regions of Sagittarius, and then affected by any chemical gradient in the progenitor. 
Nevertheless, this scenario does not seem to be capable of 
producing an analogous artificial bifurcation in the trailing tail, in the southern Galactic hemisphere \citep{KS12}.

On the other hand, \citet{PJ10} have shown that if Sagittarius was originally a late type disk, its intrinsic rotation may also have an important role in the 
properties of the bifurcation. Armlets released at successive pericenters are influenced by the relative orientation of the internal and 
orbital angular momentum of the progenitor, so to result in an apparent bifurcation. However, this model seems to clash with the 
observed kinematics of Sagittarius' remnant. While observations of the core are best explained by a pressure supported system, this scenario 
would predict that, as part of a disk, the disrupting remnant should display a higher degree of residual ordered motion \citep{PJ11}. 

Although the purpose of this paper is not to provide a model of Sagittarius, I have shown that the mechanism that generates feathers 
in the slow streams of GCs can work similarly in massive progenitors, and determine bifurcations in the density distribution of each armlet. 
A first difficulty posed by Sagittarius is that its bifurcations do not lie within the orbital plane; instead, they seem almost perpendicular to it \citep[e.g.,][]{BV14}.
For a spherical host potential, the intrinsic bifurcations generated by energy 
modulations studied in this paper lie within the orbital plane. However, internal rotation can also play a part.
If the maximum angle on the sky between the so called bright and faint streams is about $\omega_{\rm Sag}$, 
eqn.~(\ref{oplanetilt}) can be used to estimate, in order of magnitude, the degree of ordered rotation needed at escape:
 \begin{equation}
\Delta\langle\delta v_z\rangle\equiv  \langle\delta v_z\rangle_{\rm bright}-\langle\delta v_z\rangle_{\rm faint} \approx v_{{\rm p},\varphi} \arctan \omega_{\rm Sag} \ .
 \label{sagrot}
 \end{equation}
Assuming $\vp\approx220$ kms$^{-1}$ and $\omega_{\rm Sag}\approx10$ deg, I get $\Delta\langle\delta v_z\rangle\approx40$ kms$^{-1}$,
which is a very reasonable requirement for (twice) the required rotational velocity in the progenitor.

Additionally, a second important question that remains to be answered is whether Sagittarius could in fact shed a stream
that is coherent enough for the bifurcation to survive. The length of the tails, the size of the remnant's core and the number of GCs 
associated with it all testify that the total mass of Sagittarius at infall had to be considerable, most likely above $10^9\ M_{\odot}$. 
However, as it is difficult to constrain it through kinematic measurements of the remnant's core \citep[e.g.][]{PJ11}, such mass is not 
known in detail. Using Fig.~\ref{cohef}, we see that a dwarf with a mass between $10^9$ and $10^{10}\ M_{\odot}$ is in the coherent 
regime as long as $\sigma_{\rm s}\lesssim10$ kms$^{-1}$, which would allow bifurcations to survive random motions. As noted 
previously, a lower kinematic spread of escaping stars can be achieved by varying the dimensionless, structural properties of the progenitor, 
starting with the relative properties of the stellar component and of the dark matter halo of Sagittarius. 
As from the discussion of Sect.~2.2, more deeply embedded stars and/or amore extended dark halo are especially helpful.



\section{Conclusions}

This paper presents a simple analytic framework capable of explaining the main global properties of a collisionless stream.
According to three different inequalities, streams are found to be either: (i) slow or fast, (ii) coherent or hot, (iii) thin or wide.
The influence of different physical ingredients on these orderings is analysed. While first and third dichotomies are found directly 
dependent on the progenitor-to-host mass ratio, internal coherence of a stream is found to be a result of the balance between mass and 
escape conditions, so that it is more strongly affected by the internal structure of the progenitor and its characteristic size.
Scale lengths and mass are correlated, so that, simplifying, streams loose coherence when $\rh\gtrsim\rt$, while the transition 
between thin streams and umbrellas/shells happens for even more massive systems, when $\rh\gtrsim\rp$.
The role of the steepness of the host's density profile is analysed and found heavily influent in determining a stream's speed in 
gaining both length and width, substantially accelerating phase space mixing.
Differential streaming in harmonic cores is critically hampered.

Development of internal substructure is a natural outcome of the evolution of a coherent stream, in which internal mixing
is limited due to the cold escape conditions. Slow coherent streams determine the emergence of feathers, while fast coherent streams display
long bifurcated arms. While substructure had been identified as a result of the epicyclic nature of approximately circular orbits, 
in the general case of eccentric orbits the dominant mechanism is provided by a modulation in the mechanical energy of escaping particles.
Overdensities are associated to the local minima of the relative streaming speed between particles.
Purely epicyclic overdensities are made of ever-changing particles that lie temporarily closer to each other around fixed locations because
of their oscillating relative speed. For non circular progenitor's orbits, particles released at different times 
stream away from each other in a secular way. As particles released at pericenter are faster, the
streakline of each armlet (i.e. material shed between two apocenters) is folded along most of its length. In turn, particles released away from 
pericenter are more closely packed, resulting in `drifting' overdensities.  
The sizes of feathers are found directly dependent on the properties of the host potential, which motivates an effort 
towards techniques that are capable of modelling the full 6D distribution of a tidal stream, without having to rely on averaged `stream tracks'.

This paper introduces a flexible model to quickly generate tidal features that orbit within { static} spherically symmetric potentials 
\citep[see e.g., ][for a discussion of the more general case of evolving potentials]{PJ13}.
With respect to previous implementations, 
the shedding history of the progenitor can be varied at will. Also, by varying the distribution functions of the phase space coordinates at escape, 
it is possible to mimic the disruption of progenitors over a much wider mass range, covering all possible physical regimes
and associated morphologies. The major drawback of this generative model lies in the spherically symmetry of the host potential. However, this 
choice comes with the perk of cutting all `on the fly' numerical expense. The generation and evolution of a tidal 
streamer is obtained in a single shot. As this does not involve the numerical solution of the differential equations of motion,
evolution to any time is equivalent in terms of computational cost. All necessary information is conveniently stored in purposely 
optimised orbit libraries. As a consequence, this generative 
approach is (i) efficient enough to be used to fully explore the wide parameter space, (ii) flexible enough 
to model the phase space probability distribution of a stream as directly probed by the available data. 
Such a statistical framework and its performance in fitting for the potential's properties will be tested in a separate work.

Both analytic and numerical frameworks have been used to contextualise the properties of a few Milky Way streams.
For example, the mass of Willman~1 is estimated at $m\gtrsim10^6 M_{\odot}$, using the fact that 
multiple tidal tails similar to the overdensities expected in a coherent stream have been observed around its remnant core. 
Furthermore, the parallel streaks of the Anticenter stream can also be caused by internal coherence, and do necessarily 
require group infall or a substructured progenitor.
It is also interesting to consider the possibility that the bifurcation of Sagittarius' tails is similarly caused by the intrinsic bimodality
of the armlets of a coherent stream. The attractiveness of this explanation is that it can incorporate the best features 
of different previously proposed scenarios.
\begin{itemize}
\item{As in the mechanism proposed by \citet{Fe06}, it could account for the different chemical 
compositions of the two density peaks. Though shed around the same pericentric passage, the faint (bright) folding is composed 
of material escaped before (after) the first pericenter, henceforth probing outer (inner) regions of the progenitor.}
\item{As in \citet{PJ10}, in order to allow the streakline to loose its planar nature
and depart from the orbital plane, the progenitor needs internal rotation, with a magnitude that is found realistic for a dwarf galaxy 
(eqn.~(\ref{sagrot})). Differently from \citet{PJ10}, rotation would be 
necessary during the first pericentric passage only, as the bifurcations provided by the intrinsic folding of each arm would be in 
place as soon as the first pair of armlets are shed. These collect the most energetic material in the progenitor, lying at large radii as part of a 
rotationally supported disk ($v/\sigma>1$).  There is no need for the core of Sagittarius to also be rotationally supported, 
evading the constraints of \citep{PJ11}. In this scenario, Sagittarius would be structurally similar to the WLM dwarf Irregular
\citep{Le12}, in which rotation is important in the outer parts of its thick stellar disk, but central regions are in fact kinematically hot
and $v/\sigma\lesssim1$. }
\end{itemize}
Detailed analyses on this topic are deferred to a dedicated work.

\section*{Acknowledgments}
It is a pleasure to thank Vasily Belokurov for insightful comments about Sagittarius,
Denis Erkal for spotting an error in the scaling of the tidal radius in the original manuscript, 
Wyn Evans for enlightening discussion on the properties of disky dwarfs, 
and the anonymous referee for his constructive review. I also thank the Aspen Center for Physics, 
NSF Grant $\#$1066293, for their kind hospitality during the development 
of part of this paper. The Dark Cosmology Centre is funded by the DNRF.




\begin{thebibliography}{99}

\bibitem[Amorisco et al.(2014)]{AN14a} Amorisco, N.~C., Evans, N.~W., \& van de Ven, G.\ 2014, \nat, 507, 335 

\bibitem[Arnaboldi et al.(2012)]{AM12} Arnaboldi, M., Ventimiglia, G., Iodice, E., Gerhard, O., \& Coccato, L.\ 2012, \aap, 545, A37 

\bibitem[Bailer-Jones et al.(2013)]{BJ13} Bailer-Jones, C.~A.~L., Andrae, R., Arcay, B., et al.\ 2013, \aap, 559, A74 

\bibitem[Belokurov et al.(2006a)]{BV06a} Belokurov, V., Evans, N.~W., Irwin, M.~J., Hewett, P.~C., \& Wilkinson, M.~I.\ 2006, \apjl, 637, L29 
\bibitem[Belokurov et al.(2006b)]{BV06b} Belokurov, V., Zucker, D.~B., Evans, N.~W., et al.\ 2006, \apjl, 642, L137 
\bibitem[Belokurov et al.(2006c)]{BV06c} Belokurov, V., Zucker, D.~B., Evans, N.~W., et al.\ 2006, \apjl, 647, L111 

\bibitem[Belokurov et al.(2014)]{BV14} Belokurov, V., Koposov, S.~E., Evans, N.~W., et al.\ 2014, \mnras, 437, 116 

\bibitem[Besla et al.(2010)]{BG10} Besla, G., Kallivayalil, N., Hernquist, L., et al.\ 2010, \apjl, 721, L97 

\bibitem[Binney(2008)]{BJ08} Binney, J.\ 2008, \mnras, 386, L47 

\bibitem[Bonaca et al.(2014)]{BA14} Bonaca, A., Geha, M., Kuepper, A.~H.~W., et al.\ 2014, arXiv:1406.6063 
\bibitem[Bovy(2014)]{BJ14} Bovy, J.\ 2014, arXiv:1401.2985 
\bibitem[Bowden et al.(2014)]{BE14} Bowden, A., Evans, N.~W., \& Belokurov, V.\ 2014, arXiv:1409.1791 

\bibitem[Capuzzo Dolcetta et al.(2005)]{CD05} Capuzzo Dolcetta, R., Di Matteo, P., \& Miocchi, P.\ 2005, \aj, 129, 1906 

\bibitem[Carlin et al.(2010)]{Ca10} Carlin, J.~L., Casetti-Dinescu, D.~I., Grillmair, C.~J., Majewski, S.~R., \& Girard, T.~M.\ 2010, \apj, 725, 2290 
\bibitem[Carlin et al.(2012)]{Cas12} Carlin, J.~L., Majewski, S.~R., Casetti-Dinescu, D.~I., et al.\ 2012, \apj, 744, 25 

\bibitem[Carlberg et al.(2012)]{Ca12} Carlberg, R.~G., Grillmair, C.~J., \& Hetherington, N.\ 2012, \apj, 760, 75 
\bibitem[Carlberg \& Grillmair(2013)]{Ca13} Carlberg, R.~G., \& Grillmair, C.~J.\ 2013, \apj, 768, 171 %
\bibitem[Carlberg(2014)]{CR14} Carlberg, R.~G.\ 2014, arXiv:1412.2405 

\bibitem[Chou et al.(2007)]{Ch07} Chou, M.-Y., Majewski, S.~R., Cunha, K., et al.\ 2007, \apj, 670, 346 
\bibitem[Coccato et al.(2013)]{Co13} Coccato, L., Arnaboldi, M., \& Gerhard, O.\ 2013, \mnras, 436, 1322 

\bibitem[Deg \& Widrow(2013)]{De13} Deg, N., \& Widrow, L.\ 2013, \mnras, 428, 912 
\bibitem[Diaz \& Bekki(2012)]{DJ12} Diaz, J.~D., \& Bekki, K.\ 2012, \apj, 750, 36 
\bibitem[D'Onghia et al.(2010)]{DE10} D'Onghia, E., Vogelsberger, M., Faucher-Giguere, C.-A., \& Hernquist, L.\ 2010, \apj, 725, 353 

\bibitem[Erkal \& Belokurov(2014)]{DE14} Erkal, D., \& Belokurov, V.\ 2014, arXiv:1412.6035 

\bibitem[Fardal et al.(2014)]{FM14} Fardal, M.~A., Huang, S., \& Weinberg, M.~D.\ 2014, arXiv:1410.1861 
\bibitem[Fellhauer et al.(2006)]{Fe06} Fellhauer, M., Belokurov, V., Evans, N.~W., et al.\ 2006, \apj, 651, 167 

\bibitem[Foster et al.(2014)]{FC14} Foster, C., Lux, H., Romanowsky, A.~J., et al.\ 2014, \mnras, 442, 3544 
\bibitem[Fukushige \& Heggie(2000)]{TF00} Fukushige, T., \& Heggie, D.~C.\ 2000, \mnras, 318, 753 


\bibitem[Gibbons et al.(2014)]{GS14} Gibbons, S.~L.~J., Belokurov, V., \& Evans, N.~W.\ 2014, arXiv:1406.2243 

\bibitem[Grillmair \& Dionatos(2006)]{Gr06a} Grillmair, C.~J., \& Dionatos, O.\ 2006, \apjl, 643, L17 
\bibitem[Grillmair \& Johnson(2006)]{Gr06b} Grillmair, C.~J., \& Johnson, R.\ 2006, \apjl, 639, L17 
\bibitem[Grillmair(2006)]{Gr06c} Grillmair, C.~J.\ 2006, \apjl, 651, L29 
\bibitem[Grillmair \& Dionatos(2006)]{Gr06d} Grillmair, C.~J., \& Dionatos, O.\ 2006, \apjl, 641, L37 

\bibitem[Helmi \& White(1999)]{HA99} Helmi, A., \& White, S.~D.~M.\ 1999, \mnras, 307, 495 


\bibitem[Johnston(1998)]{JK98} Johnston, K.~V.\ 1998, \apj, 495, 297 
\bibitem[Johnston et al.(2001)]{JK01} Johnston, K.~V., Sackett, P.~D., \& Bullock, J.~S.\ 2001, \apj, 557, 137 
\bibitem[Johnston et al.(2008)]{JK08} Johnston, K.~V., Bullock, J.~S., Sharma, S., et al.\ 2008, \apj, 689, 936 

\bibitem[Just et al.(2009)]{Ju09} Just, A., Berczik, P., Petrov, M.~I., \& Ernst, A.\ 2009, \mnras, 392, 969 

\bibitem[Kirihara et al.(2014)]{Ki14} Kirihara, T., Miki, Y., \& Mori, M.\ 2014, arXiv:1408.4920 

\bibitem[Koposov et al.(2010)]{KS10} Koposov, S.~E., Rix, H.-W., \& Hogg, D.~W.\ 2010, \apj, 712, 260 
\bibitem[Koposov et al.(2012)]{KS12} Koposov, S.~E., Belokurov, V., Evans, N.~W., et al.\ 2012, \apj, 750, 80 


\bibitem[Khoperskov et al.(2014)]{Kh14} Khoperskov, S.~A., Moiseev, A.~V., Khoperskov, A.~V., \& Saburova, A.~S.\ 2014, arXiv:1404.1247 

\bibitem[Kleyna et al.(2003)]{Kl03} Kleyna, J.~T., Wilkinson, M.~I., Gilmore, G., \& Evans, N.~W.\ 2003, \apjl, 588, L21

\bibitem[K{\"u}pper et al.(2010)]{KA10} K{\"u}pper, A.~H.~W., Kroupa, P., Baumgardt, H., \& Heggie, D.~C.\ 2010, \mnras, 401, 105 
\bibitem[K{\"u}pper et al.(2012)]{KA12} K{\"u}pper, A.~H.~W., Lane, R.~R., \& Heggie, D.~C.\ 2012, \mnras, 420, 2700 

\bibitem[Law \& Majewski(2010)]{La10} Law, D.~R., \& Majewski, S.~R.\ 2010, \apj, 714, 229 
\bibitem[Leaman et al.(2012)]{Le12} Leaman, R., Venn, K.~A., Brooks, A.~M., et al.\ 2012, \apj, 750, 33 
\bibitem[Lora et al.(2013)]{LV13} Lora, V., Grebel, E.~K., S{\'a}nchez-Salcedo, F.~J., \& Just, A.\ 2013, \apj, 777, 65 
\bibitem[Lux et al.(2013)]{Lu13} Lux, H., Read, J.~I., Lake, G., \& Johnston, K.~V.\ 2013, \mnras, 436, 2386

\bibitem[Majewski et al.(2003)]{MS03} Majewski, S.~R., Skrutskie, M.~F., Weinberg, M.~D., \& Ostheimer, J.~C.\ 2003, \apj, 599, 1082 

\bibitem[Mart{\'{\i}}nez-Delgado et al.(2012)]{MD12} Mart{\'{\i}}nez-Delgado, D., Romanowsky, A.~J., Gabany, R.~J., et al.\ 2012, \apjl, 748, L24 

\bibitem[Ngan \& Carlberg(2014)]{NW14} Ngan, W.~H.~W., \& Carlberg, R.~G.\ 2014, \apj, 788, 181 
\bibitem[Newberg et al.(2009)]{Ne09} Newberg, H.~J., Yanny, B., \& Willett, B.~A.\ 2009, \apjl, 700, L61 

\bibitem[Odenkirchen et al.(2001)]{Od01} Odenkirchen, M., Grebel, E.~K., Rockosi, C.~M., et al.\ 2001, \apjl, 548, L165 
\bibitem[Odenkirchen et al.(2002)]{Od02} Odenkirchen, M., Grebel, E.~K., Dehnen, W., Rix, H.-W., \& Cudworth, K.~M.\ 2002, \aj, 124, 1497 
\bibitem[Odenkirchen et al.(2009)]{Od09} Odenkirchen, M., Grebel, E.~K., Kayser, A., Rix, H.-W., \& Dehnen, W.\ 2009, \aj, 137, 3378 


\bibitem[Pe{\~n}arrubia et al.(2010)]{PJ10} Pe{\~n}arrubia, J., Belokurov, V., Evans, N.~W., et al.\ 2010, \mnras, 408, L26 
\bibitem[Pe{\~n}arrubia et al.(2011)]{PJ11} Pe{\~n}arrubia, J., Zucker, D.~B., Irwin, M.~J., et al.\ 2011, \apjl, 727, L2 
\bibitem[Pe{\~n}arrubia et al.(2012)]{PJ12} Pe{\~n}arrubia, J., Koposov, S.~E., \& Walker, M.~G.\ 2012, \apj, 760, 2 
\bibitem[Pe{\~n}arrubia(2013)]{PJ13} Pe{\~n}arrubia, J.\ 2013, \mnras, 433, 2576 

\bibitem[Perryman et al.(2001)]{Pe01} Perryman, M.~A.~C., de Boer, K.~S., Gilmore, G., et al.\ 2001, \aap, 369, 339 
\bibitem[Price-Whelan et al.(2014)]{PW14} Price-Whelan, A.~M., Hogg, D.~W., Johnston, K.~V., \& Hendel, D.\ 2014, arXiv:1405.6721 

\bibitem[Read et al.(2006)]{RJ06} Read, J.~I., Wilkinson, M.~I., Evans, N.~W., Gilmore, G., \& Kleyna, J.~T.\ 2006, \mnras, 366, 429 

\bibitem[Romanowsky et al.(2012)]{RA12} Romanowsky, A.~J., Strader, J., Brodie, J.~P., et al.\ 2012, \apj, 748, 29 
\bibitem[Renaud et al.(2011)]{RF11} Renaud, F., Gieles, M., \& Boily, C.~M.\ 2011, \mnras, 418, 759 

\bibitem[S{\'a}nchez-Salcedo \& Lora(2010)]{SS10} S{\'a}nchez-Salcedo, F.~J., \& Lora, V.\ 2010, \mnras, 407, 1135 

\bibitem[Sanders \& Binney(2013)]{SJ13a} Sanders, J.~L., \& Binney, J.\ 2013, \mnras, 433, 1813 
\bibitem[Sanders \& Binney(2013)]{SJ13b} Sanders, J.~L., \& Binney, J.\ 2013, \mnras, 433, 1826 
\bibitem[Sanders(2014)]{SJ14} Sanders, J.~L.\ 2014, \mnras, 443, 423 
\bibitem[Sanderson et al.(2014)]{SR14} Sanderson, R.~E., Helmi, A., \& Hogg, D.~W.\ 2014, IAU Symposium, 298, 207 
\bibitem[Shi et al.(2012)]{Sh12} Shi, W.~B., Chen, Y.~Q., Carrell, K., \& Zhao, G.\ 2012, \apj, 751, 130 
\bibitem[Siegal-Gaskins \& Valluri(2008)]{SG08} Siegal-Gaskins, J.~M., \& Valluri, M.\ 2008, \apj, 681, 40 

\bibitem[Varghese et al.(2011)]{Va11} Varghese, A., Ibata, R., \& Lewis, G.~F.\ 2011, \mnras, 417, 198 
\bibitem[Veljanoski et al.(2014)]{Ve14} Veljanoski, J., Mackey, A.~D., Ferguson, A.~M.~N., et al.\ 2014, \mnras, 442, 2929 
\bibitem[Vera-Ciro \& Helmi(2013)]{VC13} Vera-Ciro, C., \& Helmi, A.\ 2013, \apjl, 773, L4 

\bibitem[Yoon et al.(2011)]{Yo11} Yoon, J.~H., Johnston, K.~V., \& Hogg, D.~W.\ 2011, \apj, 731, 58 

\bibitem[Willman et al.(2006)]{WB06} Willman, B., Masjedi, M., Hogg, D.~W., et al.\ 2006, arXiv:astro-ph/0603486 
\bibitem[Willman et al.(2011)]{WB11} Willman, B., Geha, M., Strader, J., et al.\ 2011, \aj, 142, 128 

\end{thebibliography}
\end{document}